\documentclass{bioinfo_notitles}
\copyrightyear{2016}
\pubyear{2016}

\usepackage{amssymb,amsfonts,amsmath}
\usepackage{xspace}
\usepackage[title]{appendix}
\usepackage[table]{xcolor}
\usepackage{natbib}
\usepackage{enumerate,float,indentfirst}
\usepackage{xspace}
\usepackage{multirow}
\usepackage{subfigure}
\usepackage[noend]{algpseudocode}
\usepackage{algorithm}
\usepackage[unicode=true,hidelinks]{hyperref}
\usepackage[group-separator={,}]{siunitx}

\newlength{\widecommentlength}
\newcommand{\widecommentbox}[2]{\def#1##1{\strut\newline\noindent\colorbox{#2}{\linespread{1}\parbox{.45\textwidth}{\small
##1}}\newline}}
\widecommentbox{\anton}{orange!25!white}
\widecommentbox{\yana}{green!60!white}
\widecommentbox{\sasha}{blue!25!white}
\widecommentbox{\pavel}{red!60!white}
\widecommentbox{\sergey}{violet!60!white}
\widecommentbox{\andrey}{pink!60!white}

\def\igrc {\textsc{IgReC}\xspace}
\def\igrepcon {\textsc{IgRepertoireConstructor}\xspace}
\def\vjfinder{\textsc{VJ~Finder}\xspace}

\def\igsimulator{\textsc{IgSimulator}\xspace}
\def\hgconstructor{\textsc{HG~Constructor}\xspace}
\def\bh{\textsc{BayesHammer}\xspace}
\def\hammer{\textsc{Hammer}\xspace}

\def\igblast{\textsc{IgBLAST}\xspace}

\def\mixcr{\textsc{MiXCR}\xspace}
\def\imseq{\textsc{IMSEQ}\xspace}
\def\kplus{\textsc{BlockAlignment}\xspace}
\def\migec{\textsc{MiGEC}\xspace}

\def\Reads{\textsc{Reads}}
\def\Read{\textsc{Read}}

\def\Kmers{\textsc{Kmers}}

\begin{document}

\title[\igrc]{New algorithmic challenges of adaptive immune repertoire construction}

\author[Shlemov et~al]{
      Alexander Shlemov\,$^{1,2,4}$,
      Sergey Bankevich\,$^1$,
      Andrey Bzikadze\,$^{1,2}$,
      Yana Safonova $^{1,3,4}$ \footnote{to whom correspondence should be addressed}}

\address{
$^{1}$Center for Algorithmic Biotechnology, Institute of Translational Biomedicine, St. Petersburg State University, Russia\\
$^{2}$Department of Statistical Modeling, Division of Mathematics and Mechanics, St. Petersburg State University, Russia\\
$^{3}$Algorithmic Biology Laboratory, St. Petersburg Academic University, Russia\\
$^{4}$These authors have contributed equally.
}

\history{}

\editor{}

\maketitle

\begin{abstract}
\section{Motivation: }
The analysis of antibodies and T-cell receptors (TCRs) concentrations in serum is a fundamental problem in immunoinformatics.
Repertoire construction is a preliminary step of analysis of clonal lineages,
understanding of immune response dynamics,
population analysis of immunoglobulin and TCR loci.
Emergence of MiSeq Illumina sequencing machine in 2013 opened horizons of investigation of adaptive immune repertoires using
highly accurate reads.
Reads produced by MiSeq are able to cover repertoires of moderate size.
At the same time, throughput of sequencing machines increases from year to year.
This will enable ultra deep scanning of adaptive immune repertoires and analysis of their diversity.
Such data requires both efficient and highly accurate repertoire construction tools.
In 2015 \cite{Safonova2015igrc} presented \igrepcon, a tool for accurate construction of antibody repertoire and immunoproteogenomics analysis.
Unfortunately, proposed algorithm was very time and memory consuming and could be a bottleneck of processing large immunosequencing libraries.
In this paper we overcome this challenge and present \igrc, a novel algorithm for adaptive repertoire construction problem.
\igrc reconstructs a repertoire with high precision even if each input read contains sequencing errors and performs well on contemporary datasets.
Results of computational experiments show that \igrc improves state-of-the-art in the field.

\section{Availability: }
\igrc is an open source and freely available program running on Linux platforms.
The source code is available at GitHub: \url{yana-safonova.github.io/ig_repertoire_constructor}.

\section{Contact: } \href{mailto:safonova.yana@gmail.com}{safonova.yana@gmail.com}

\end{abstract}

\section{Introduction}
\begin{figure}
  \begin{center}
    \includegraphics[width = .45\textwidth]{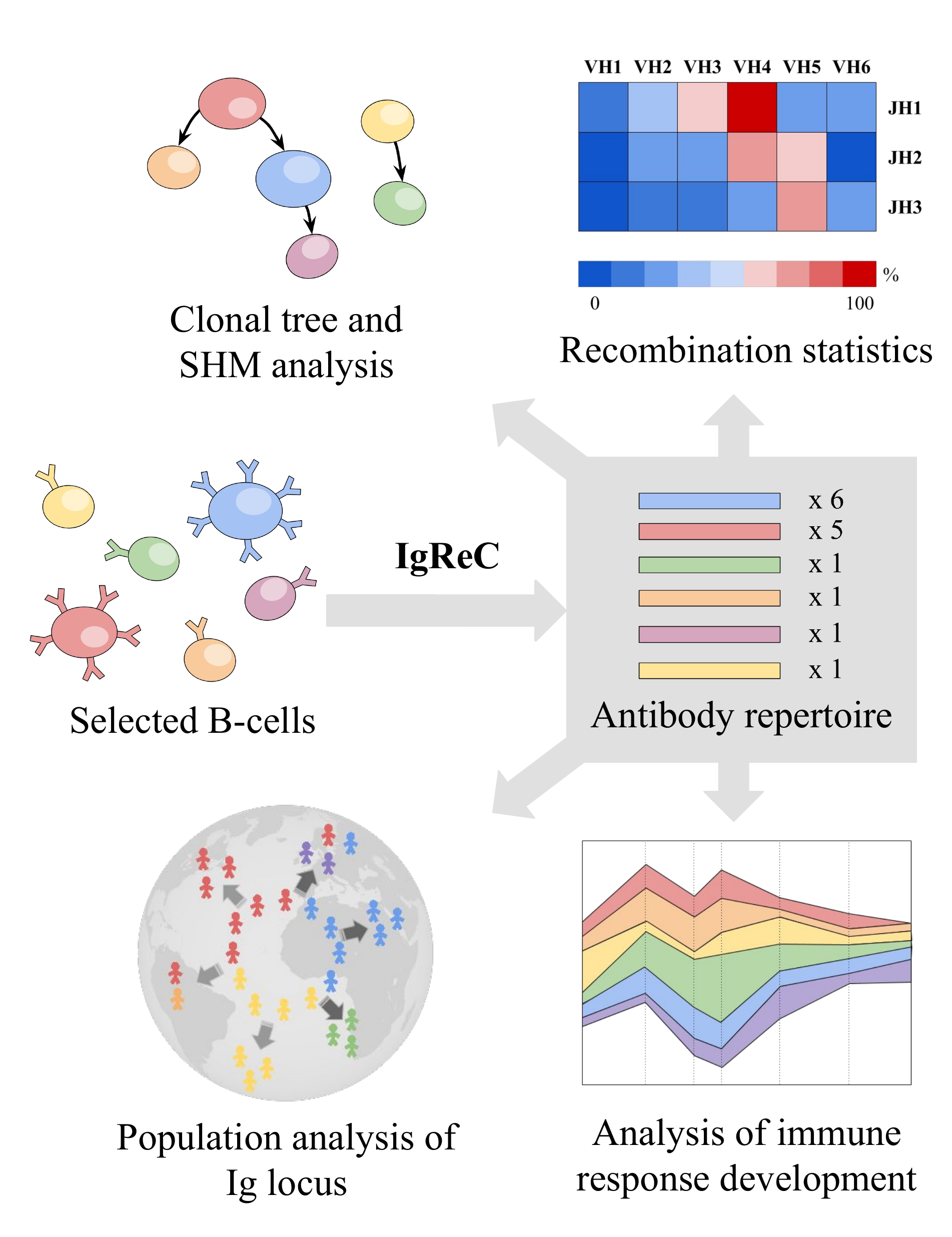}
  \end{center}
  \caption{Highly accurate antibody repertoire constructed by \igrc can be used as an input for clonal analysis and analysis of somatic hypermutations, analysis of recombination events, population analysis of immunoglobulin locus and detection of novel alleles, analysis of immune response development and treatment monitoring.}
  \label{fig:immunogenomics}
\end{figure}

Rapid development of sequencing technologies enables deep full-length scanning of adaptive immune repertoires and opens new immunoinformatics challenges (see recent reviews by \cite{georgiou2014promise}, \cite{Robinson2015}, \cite{Yaari2015} and \cite{Grieff2015}).
High accuracy of BCR and TCR repertoires allows one to solve such immunological problems as (Fig.~\ref{fig:immunogenomics}):
analysis of clonal lineages (\cite{Barak2008, Gupta2015}),
statistical analysis of recombination events and secondary diversification (\cite{Murugan2012, Elhanati2015}),
analysis of development of immune response (\cite{Bolotin2012, Laserson2014}),
population analysis of immunoglobulin and TCR loci (\cite{GadalaMaria2015}).

Construction of full-length repertoires is a different problem than the well-studied
VDJ classification (\cite{IgBlast2013}; \cite{gaeta2007ihmmune}; \cite{Elhanati2015}; \cite{Bonissone2015})
and CRD3 classification (\cite{Robins2009, Freeman2009, Robins2010, Warren2011}) problems.
In fact, VDJ classification, CDR3 classification and full-length classification are three different clustering problems with increasing granularity of partition into clusters and different biological applications.
Until 2013, there were few attempts to perform full-length clustering since it was nearly impossible to derive an accurate repertoire with previous experimental approaches based on error-prone and low coverage 454 sequencing technology.

Emergence of MiSeq Illumina sequencing machine in 2013 opened horizons of adaptive immune repertoires investigation using highly accurate $250 \times 2$ Illumina reads.
Availability of Illumina MiSeq reads raised interest of bioinformaticians to the problem and in 2015 three new tools for construction of immune repertoire from sequencing data were released: \igrepcon by \cite{Safonova2015igrc}, \mixcr by \cite{Bolotin2015} and \imseq by \cite{Kuchenbecker2015}.
Note that although the total number of B-cells in human organism can be estimated,
Illumina MiSeq does not allow one to capture its overall diversity.
Particularly, a number of different clones and their typical abundances are still unknown.

In 2015 Illumina released a new kit for HiSeq sequencing machine producing hundreds of millions $250 \times 2$ reads per run.
Such ultra high-throughput sequencing will allow one to identify even lowly abundant clones and would provide an insight into the diversity of the whole repertoire.
On the other hand ultra high-throughput sequencing needs very efficient algorithms to process.
In particular this will impose a challenge for repertoire construction tools.

Another challenge is that even modern Illumina technologies produce imperfect reads containing random errors.
Moreover, standard protocols of immunosequencing sample preparation include amplification stage.
Amplification process produces copies of RNA molecules, but introduces erroneous substitutions and short indels (\cite{Pienaar2006}).
These errors can be copied by further amplification cycles and thus mixed up with natural variations.
Abundance of sequencing and amplification errors force one to perform preliminary error correction step to be able to
construct repertoire of \textit{true antibody or TCR sequences}.

In 2015 \cite{Safonova2015igrc} released \igrepcon: an algorithm for construction of antibody repertoire from
Illumina MiSeq reads and immunoproteogenomics analysis.
\igrepcon considers unique features of adaptive immune repertoires and efficiently corrects sequencing and amplification errors
preserving significant part of natural diversity.
\igrepcon reconstructs repertoire with high precision even if each input read contains sequencing errors.
Thus, \igrepcon is able to recover repertoire even from error-prone reads.
Unfortunately, repertoire construction algorithm proposed in \cite{Safonova2015igrc} was very time and memory consuming that
could be a bottleneck in processing large immunosequencing libraries.

In this paper we overcome this challenge and present \igrc, a novel efficient algorithm for immune repertoire construction from sequencing reads.
We benchmark \igrc against recently released adaptive immune repertoires construction tools, namely \mixcr by \cite{Bolotin2015} and \imseq by \cite{Kuchenbecker2015}.
For benchmarking we use datasets simulated using \igsimulator tool by \cite{igsimulator2015} and show high accuracy of \igrc results compared to other solutions (see Results).
We also analyze the results of \igrc on immunosequencing data from sorted B-cells with various diversity and expression levels (naive, antibody-secreting and plasma cells) and show that \igrc reflects unique features of different B-cell subtypes.
The results of computational experiments show that \igrc improves state-of-the-art of adaptive immune repertoire construction problem.

\section{Methods}

\subsection{\igrc pipeline}
\igrc takes immunosequencing reads covering entire variable regions of immunoglobulins as an input and
computes antibody repertoire as a partition of reads into clusters.
Each cluster presents unique antibody and is characterized by its sequence and multiplicity.
Multiplicity of a cluster is computed as the number of reads that form it.
Sequence of a cluster is consensus between the reads.

Ideally, clusters are formed by identical reads.
However, in fact reads corresponding to the same cluster can be similar, but not identical due to sequencing and amplification errors.
We define similar reads as ones with Hamming distance small enough
(in \cite{Safonova2015igrc} we introduced definition of Hamming distance for sequences of non-equal length).
We use Hamming graph constructed on reads and expect that reads corresponding to identical antibodies form dense structures in this graph.

\begin{figure}
  \begin{center}
    \includegraphics[width = .4\textwidth]{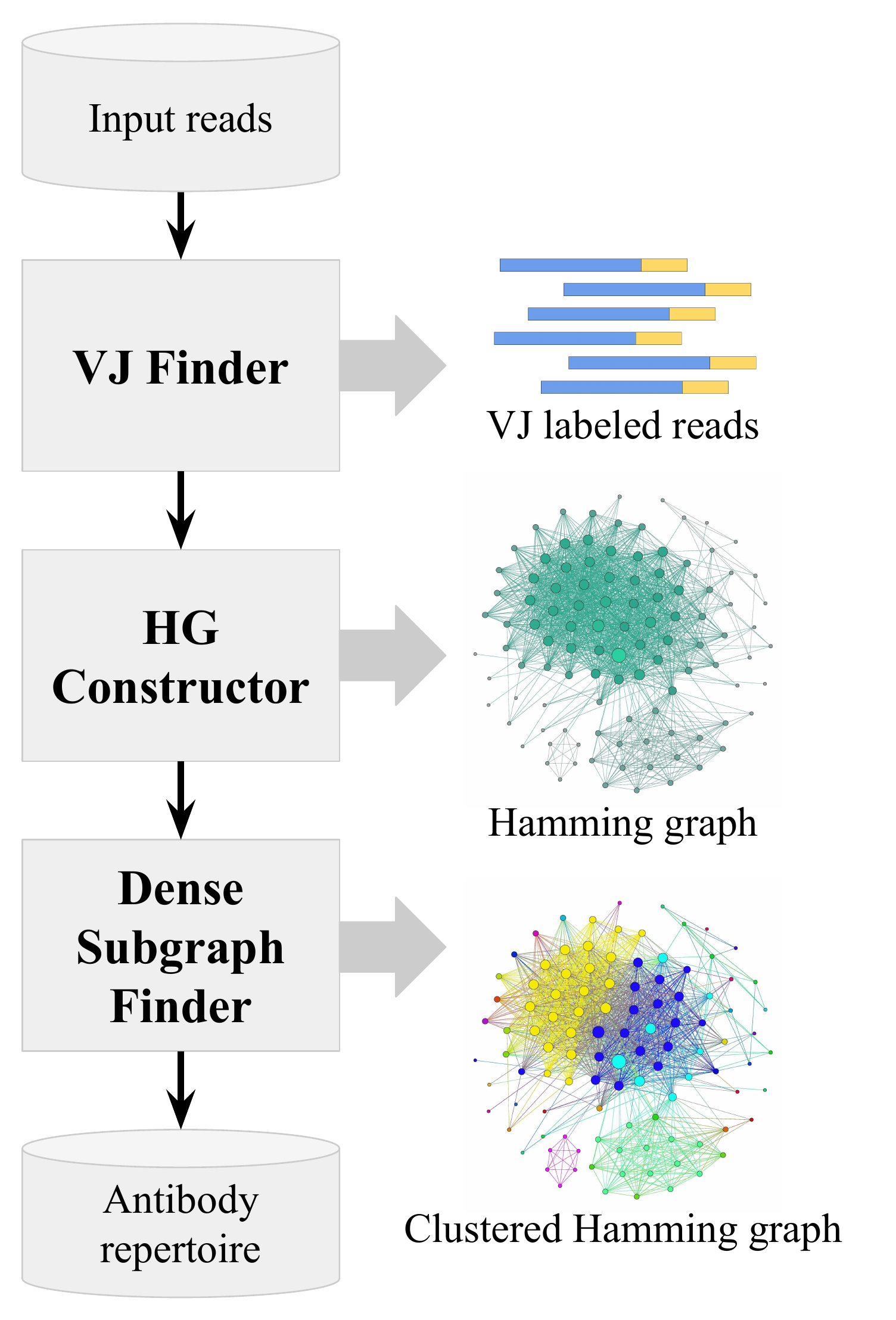}
  \end{center}
  \caption{\label{fig:igrc_pipeline}
    \igrc pipeline includes \vjfinder, \hgconstructor and \textsc{Dense~Subgraph~Finder} steps.}
\end{figure}

Fig.~\ref{fig:igrc_pipeline} shows \igrc pipeline consisting of the following steps:
\vjfinder, \hgconstructor and \textsc{Dense~Subgraph~Finder}.

Immunosequencing library may contain contaminated reads, i.e., reads irrelevant to immunoglobulins.
Thus, the first step of \igrc (\vjfinder) filters out irrelevant reads using alignment against database of Ig germline genes.
Otherwise, those reads increase amount of computations and prevent effective read clustering.

Then \hgconstructor builds Hamming graph on Ig-seq reads.
The graph uses reads as vertices and has an edge between two vertices if Hamming distance between corresponding reads does not exceed predefined threshold $\tau$.

Finally, \textsc{Dense~Subgraph~Finder} computes clusters of the constructed Hamming graph which correspond to groups of similar reads.

Further we describe \vjfinder and \hgconstructor algorithms.
Algorithm of \textsc{Dense~Subgraph~Finder} and construction of final read clusters were described in \cite{Safonova2015igrc}.

\subsection{\vjfinder}
\vjfinder is the first step of \igrc that performs filtering of contaminated reads using alignment against Ig germline genes (we use IMGT database by \cite{lefranc2009imgt}).
Also \vjfinder filters out reads that do not fully cover V(D)J region since their error correction can result in false inferences.

Alignment of immunosequencing reads against germline database or \emph{V(D)J labeling} is well-known problem \cite[]{IgBlast2013,gaeta2007ihmmune,Elhanati2015,Bonissone2015}.
However, solutions provided by existing approaches are excessive for our purposes: 
to distinguish immunosequencing reads from contaminations we do not need to identify all recombination events.
It is sufficient to detect presence of both V and J gene segments.
Thus, we propose an ultra-fast alignment algorithm to filter reads corresponding to full-length antibody sequences.

\begin{figure*}
  \begin{center}
    \mbox{
      \subfigure[]{
          \includegraphics[width=.35\textwidth]{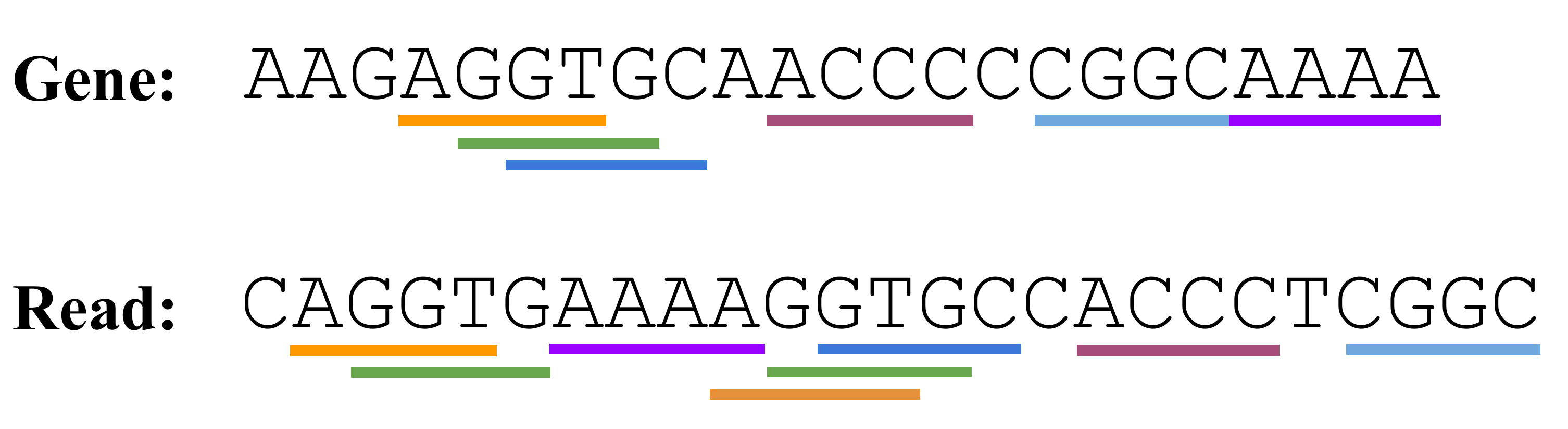}
            \label{fig:kmer_matches}
      }
      \hspace{.1\textwidth}
      \subfigure[]{
        \includegraphics[width=.35\textwidth]{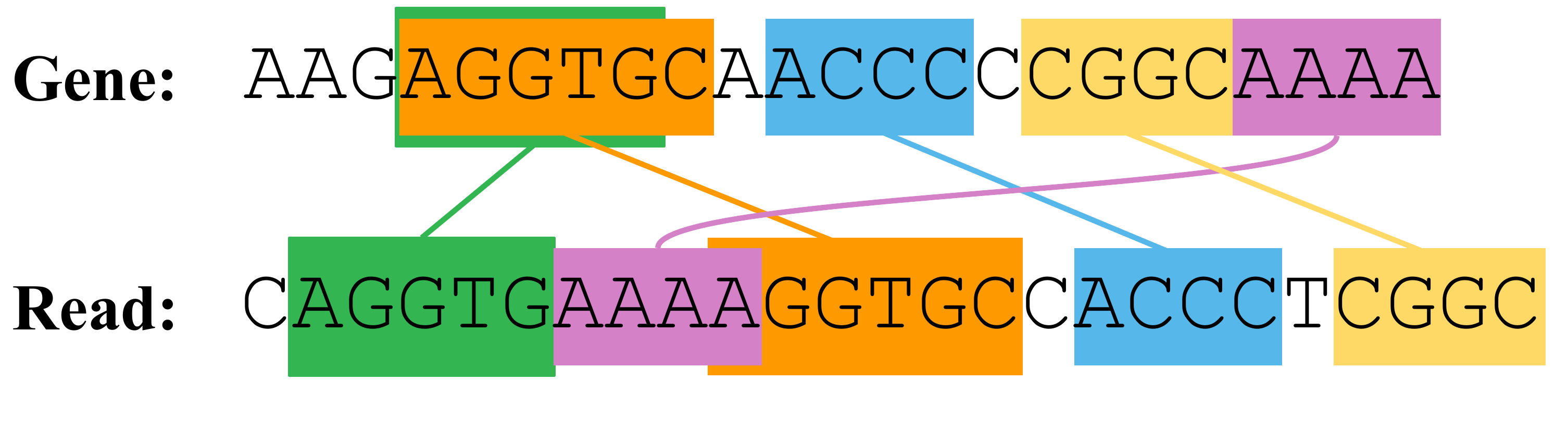}
        \label{fig:k_plus_matches}
      }
    }\\
    \mbox{
      \subfigure[]{
          \includegraphics[width=.35\textwidth]{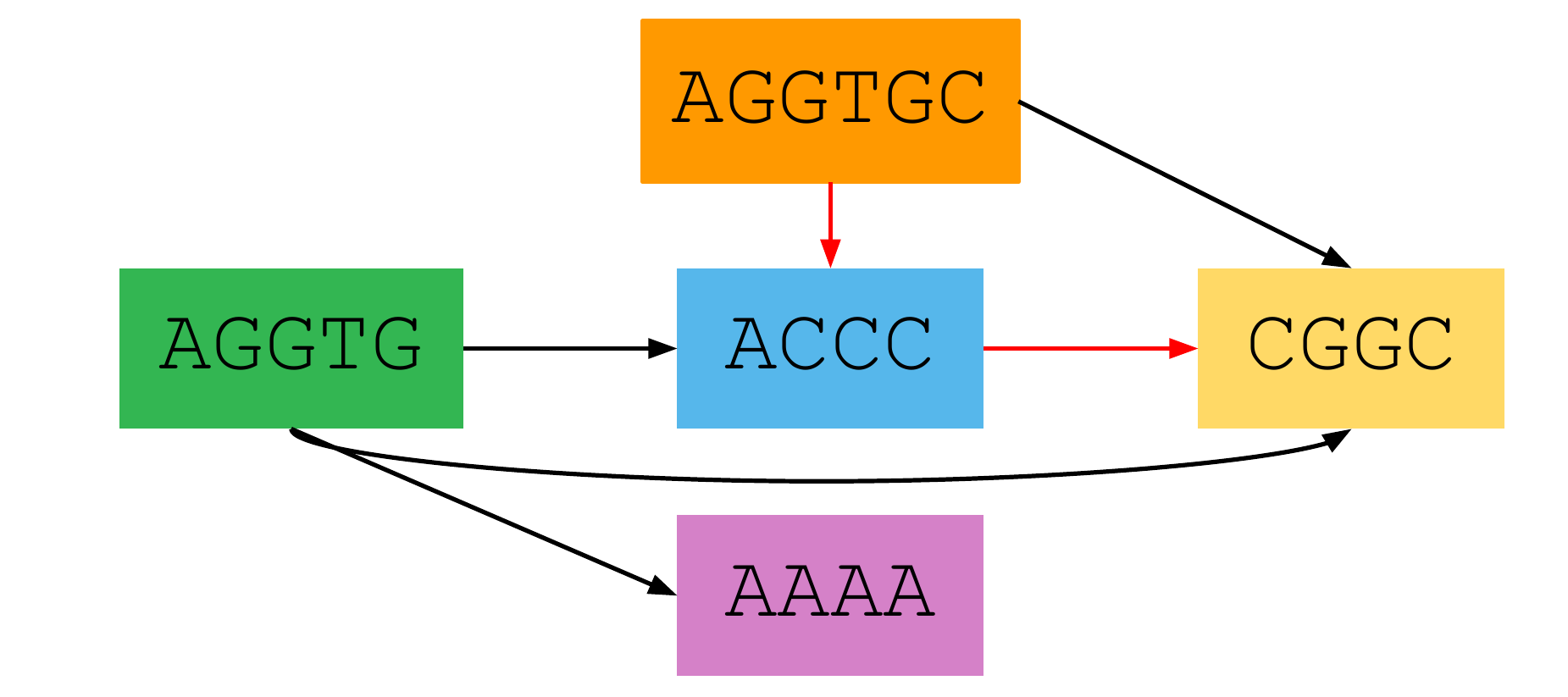}
            \label{fig:consistency_graph}
      }
      \hspace{.1\textwidth}
      \subfigure[]{
        \includegraphics[width=.35\textwidth]{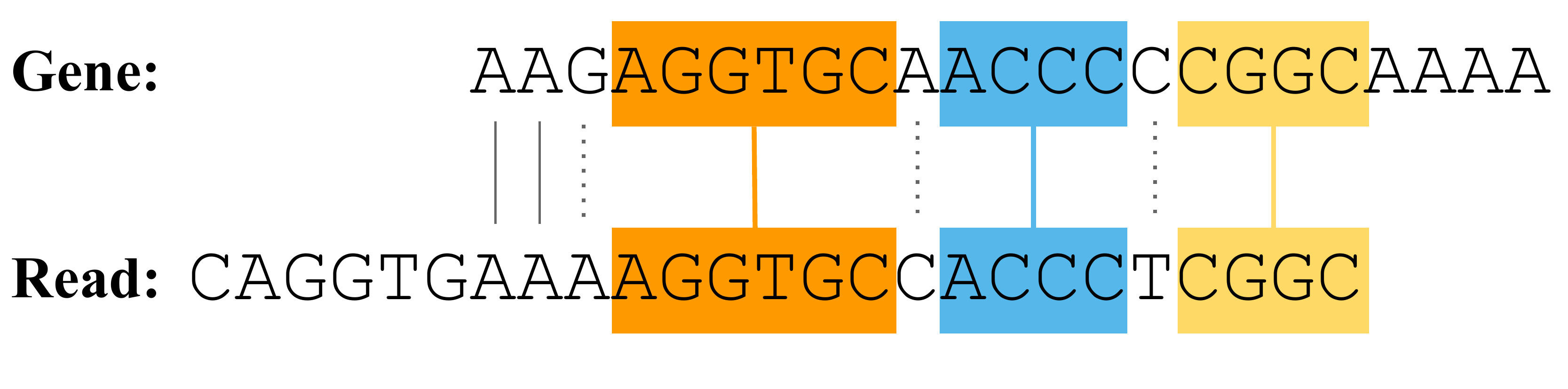}
        \label{fig:final_alignment}
      }
    }
  \end{center}
  \caption{
  \label{fig:kplus_algorithm}
  Example of \kplus algorithm work.\newline
  (Upper left) On the first step algorithm searches for $k$ matches: $k$-mers shared between read and Ig germline gene (in our case, $k = 4$).
  Each $k$ match is characterized by a pair $(r, g)$, where $r$ and $g$ are $k$-mer positions on read and gene, respectively.\newline
  (Upper right) If $k$-mers are consecutive in both read and gene, \kplus algorithm joins them and constructs blocks.
  Each block is characterized by $(r, g, l)$, where $r$ and $g$ are positions on read and gene and $l$ is a length of resulting block ($l \geq k$).
  For example, orange (\textsc{AGGT}) , green (\textsc{GGTG}) and blue (\textsc{GTGC}) $k$-mers are consecutive in both read and gene and thus will be joined into block \textsc{AGGTGC} of length $6$.
  Note that blocks can be overlapping (e.g., green block \textsc{AGGTG} is a prefix of orange block \textsc{AGGTG} on read, but both of them have different positions on gene).\newline
  (Lower left) At the third step, \kplus algorithm constructs consistency graph on blocks.
  Two blocks $(r_1, g_1, l_1)$ and $(r_2, g_2, l_2)$ are adjacent in the consistency graph if they are consistently ordered, i.e., $r_1< r_2$ and $g_1 < g_2$.
  For example, this condition is not satisfied for orange (\textsc{AGGTGC}) and pink (\textsc{AAAA}) blocks, thus corresponding edge is not presented in consistency graph.
  \kplus algorithm computes a path with the best score, where score is the total number of matching positions.
  Path with the best score $14$ consists of blocks \textsc{AGGTGC}, \textsc{ACCC} and \textsc{CGGC} (corresponding edges are highlighted in red).\newline
  (Lower right) Any path in the consistency graph can be converted into an alignment between read and gene.
  The score of the constructed alignment is computed as a number of matching positions (in our case, alignment score is $16$).
  }
\end{figure*}

\paragraph{\vjfinder pipeline.}
\vjfinder algorithm computes the best V and J hits for each input read using score computed by \kplus algorithm.
\kplus algorithm computes local alignment between immunosequencing read \textsc{Read} and Ig germline gene \textsc{Gene} (see pseudocode in Algorithm \ref{alg:kplusaignment}).
Fig.~\ref{fig:kplus_algorithm} illustrates details of \kplus algorithm.

\textsc{FindSharedKmers} procedure (Fig.~\ref{fig:kmer_matches}) identifies positions of $k$-mers shared between \textsc{Read} and \textsc{Gene}.
Result of this procedure is a set of \textit{$k$ matches}, i.e., pairs $(r, g)$, where
$r$ and $g$ are start positions of $k$-mer in \textsc{Read} and \textsc{Gene}, respectively.

\textsc{JoinConsequentKmers} procedure (Fig.~\ref{fig:k_plus_matches}) joins consecutive $k$ matches to \textit{blocks}.
Each block presents a triple $(r, g, l)$, where $l$ is a length of the resulting block.

\textsc{ConstructConsistencyGraph} procedure (Fig~\ref{fig:consistency_graph}) constructs graph on blocks using \textit{consistency rule}: two blocks $(r_1, g_1, l_1)$ and $(r_2, g_2, l_2)$ are adjacent if they are consistently ordered, i.e., $r_1 < r_2$ and $g_1 < g_2$.
The constructed consistency graph is a \emph{directed acyclic graph (DAG).}

For each path in the consistency graph we compute its \textit{path score} as the total number of computed blocks.
\textsc{FindLongestPath} procedure finds a path with the largest score in the constructed
consistency graph as a solution to the \emph{weighted longest path problem} in a DAG.

\textsc{ConstructAlignment} procedure (Fig.~\ref{fig:final_alignment}) computes an alignment by
filling gaps between the blocks in the computed longest path.
Alignment score is computed as the overall number of matching positions.

\begin{algorithm}
\caption{\kplus workflow}\label{alg:kplusaignment}
\begin{algorithmic}[1]
\Procedure{BlockAlignment}{\textsc{Read}, \textsc{Gene}, $k$}
  \State $\textsc{$k$-mers} \gets $ \Call{FindSharedKmers}{\textsc{Read}, \textsc{Gene}, $k$}
  \State $\textsc{Blocks} \gets $ \Call{JoinConsequentKmers}{\textsc{$k$-mers}, $k$}
  \State $\textsc{cGraph} \gets $ \Call{ConstructConsistencyGraph}{\textsc{Blocks}}
  \State $\textsc{Path} \gets $ \Call{FindLongestPath}{\textsc{cGraph}}
  \State $\textsc{Alignment, Score} \gets $ \Call{ConstructAlignment}{\textsc{Path}}
  \State \Return \textsc{Alignment, Score}
\EndProcedure
\end{algorithmic}
\end{algorithm}

\paragraph{\vjfinder discussion.}
Since region before V gene region and constant region after J gene segment are not significant for full-length repertoire construction,
\vjfinder crops all reads by the first position of V gene segment and the last position of J gene segment.
It simplifies significantly Hamming graph construction since cropped reads start from the first position of V gene. 

\kplus algorithm used by \vjfinder has quadratic complexity, but still is effective due to short length of input reads. 
\vjfinder beats standard VDJ labeling approaches in term of resource consumption without loss of accuracy.
We show that \vjfinder correctly aligned $\num{1405177}$ reads within $\sim 6$ min using $16$ threads vs $\sim 30$~hours of \igblast work
(Appendix~A).
Sensitivity of \vjfinder algorithm can be tuned using parameters $k_V$ ($k$-mer size for V alignment) and $k_J$ ($k$-mer size for J alignment).
By default, \vjfinder uses $k_V = 7$ and $k_J = 5$ (see discussion on $k$-mer size selection in \cite{IgBlast2013}) and correctly processes even highly mutated sequences.

\subsection{\hgconstructor}
\hgconstructor builds Hamming graph (HG) on reads reported by \vjfinder as Ig relevant.
Construction of HG is a challenging computational problem that was solved in \hammer (\cite{medvedev2011error}) and \bh
(\cite{Nikolenko2013}) algorithms for $k$-mers of genomic reads.
Specific features of immunosequencing reads prevent application of \hammer and \bh to this problem.
\hgconstructor was specially designed for immune sequences of variable length and constructs HG on reads instead of $k$-mers.
In this section we describe a novel algorithm for efficient construction of HG on immune sequences.

\paragraph{HG construction problem.}
To construct Hamming graph for a given set of strings (e.g., reads or $k$-mers) and threshold $\tau$ one needs to find all pairs $(s_1, s_2)$ such that $\mathcal{HD}(s_1, s_2) \leq \tau$, where $\mathcal{HD}$ is Hamming distance.
We use the definition of Hamming distance of two strings of different length that was introduced in \cite{Safonova2015igrc}
\footnote{After \vjfinder immunosequencing reads start with the first position of V gene. 
This allows us to compute $\mathcal{HD}$ on length of the shortest read from pair.}.

For strings of equal length $L$ and $\tau = 1$ \cite{Knuth1998} suggested an approach based on the following observation.
Strings differing only by a single mismatch must match exactly in either the left of in the right half. 
To show this lets divide each string into left and right parts (of equal length for even $L$ or almost equal length for odd $L$).
Then any two pairs of strings differing by the only mismatch must share the left or the right part and one can compute $\mathcal{HD}$ only for such pairs.
In \hammer and \bh this idea was extended for arbitrary number of mismatches.

However, these approaches are applicable for sequences of equal lengths only.
\cite{Safonova2015igrc} extended an \bh approach for reads of various lengths. 
This algorithm works well for small input but becomes prohibitively slow for large Ig-seq libraries ($>3$M reads).
Immunosequencing reads provide an additional challenge for proposed algorithms since they may share significantly long part (e.g., common V gene segment). 
In this case, an algorithm forces to compute $\mathcal{HD}$ for almost all pairs of reads sharing the same V.

\begin{figure*}
	\begin{center}
		\mbox{
			\subfigure[] {
				\includegraphics[width = .32\textwidth]{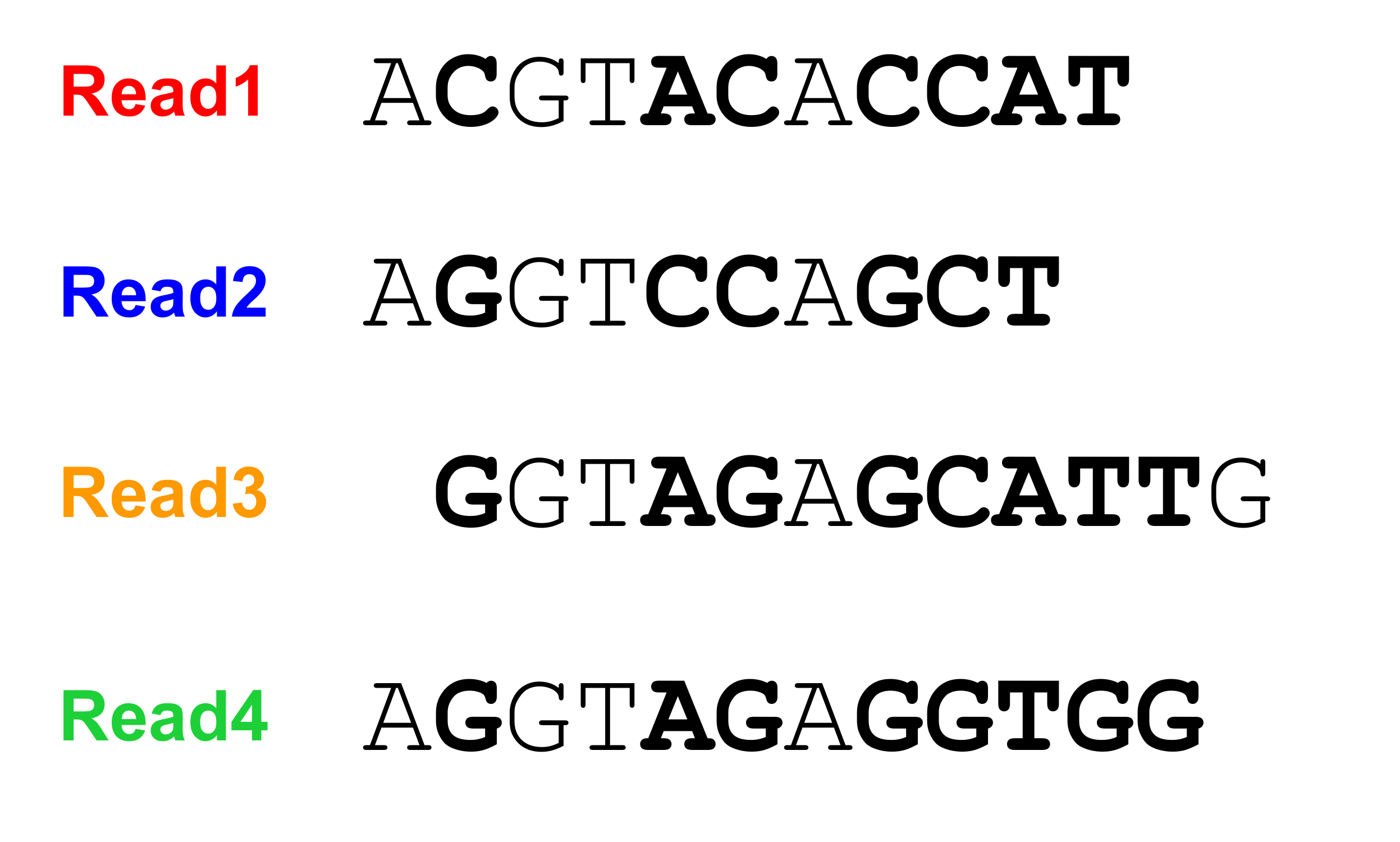}
			}
			\hspace{.05\textwidth}
			\subfigure[] {
				\includegraphics[width = .4\textwidth]{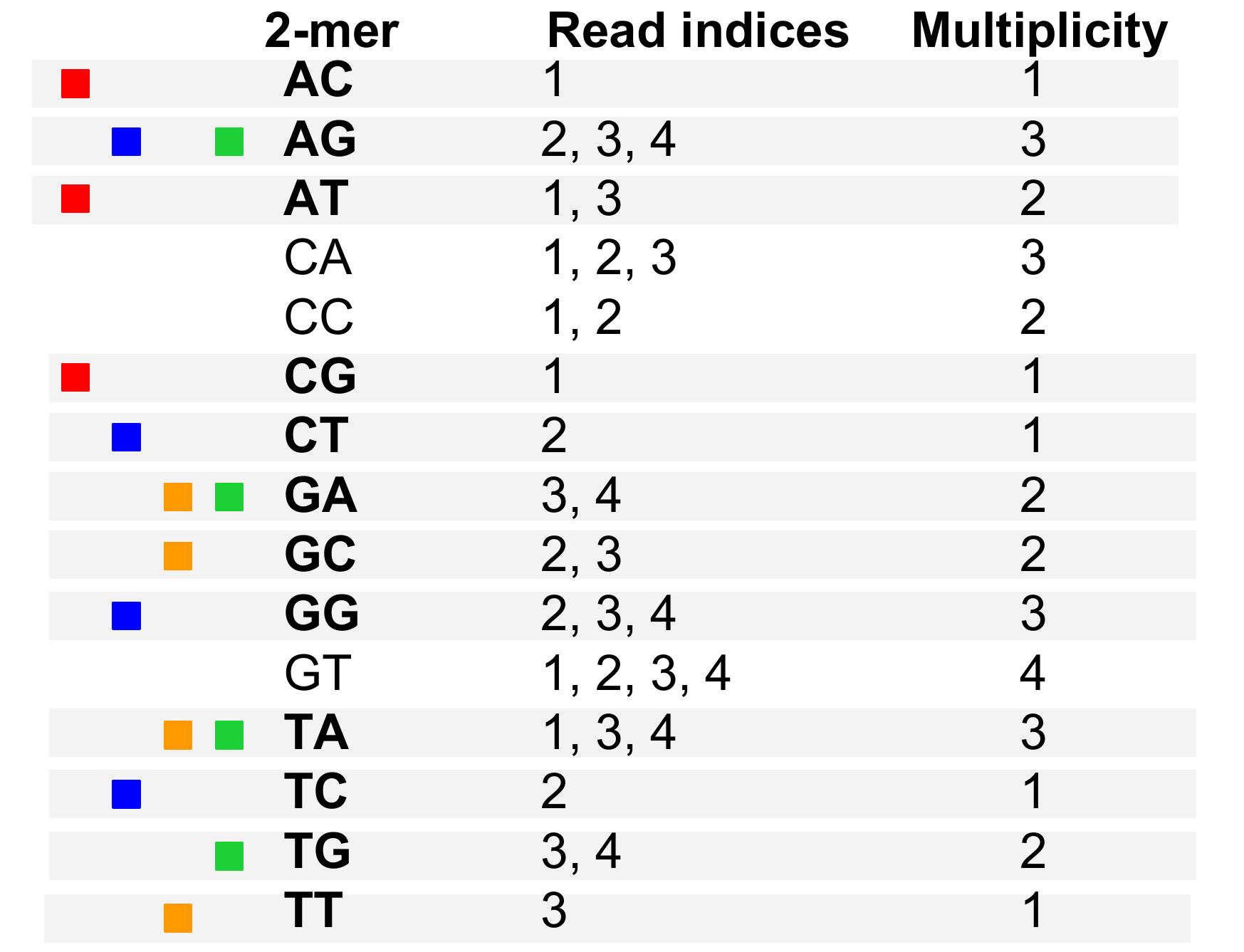}
			}			
		}\\
		\mbox{
			\subfigure[] {
				\includegraphics[width = .32\textwidth]{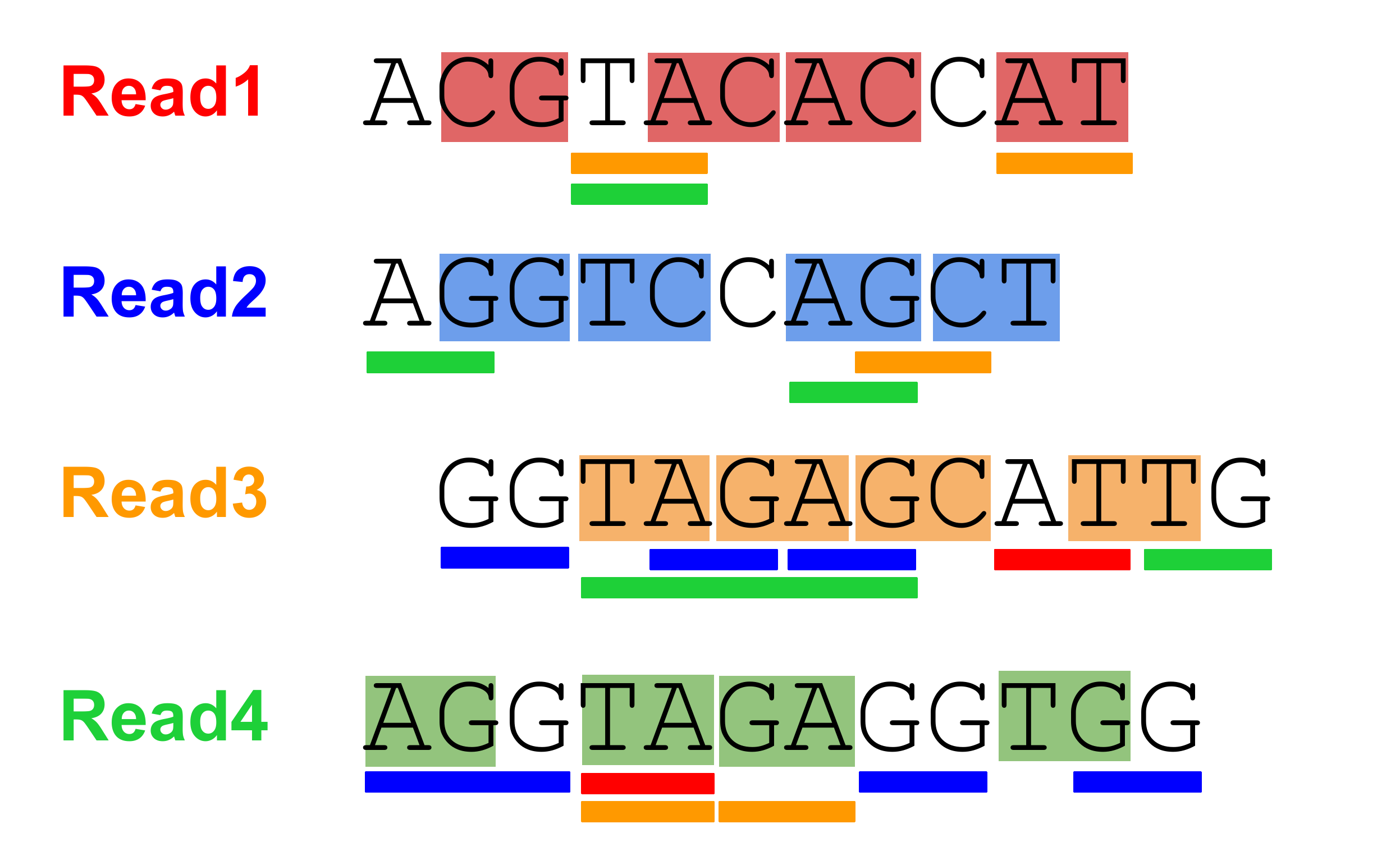}
			}
			\subfigure[] {
				\includegraphics[width = .45\textwidth]{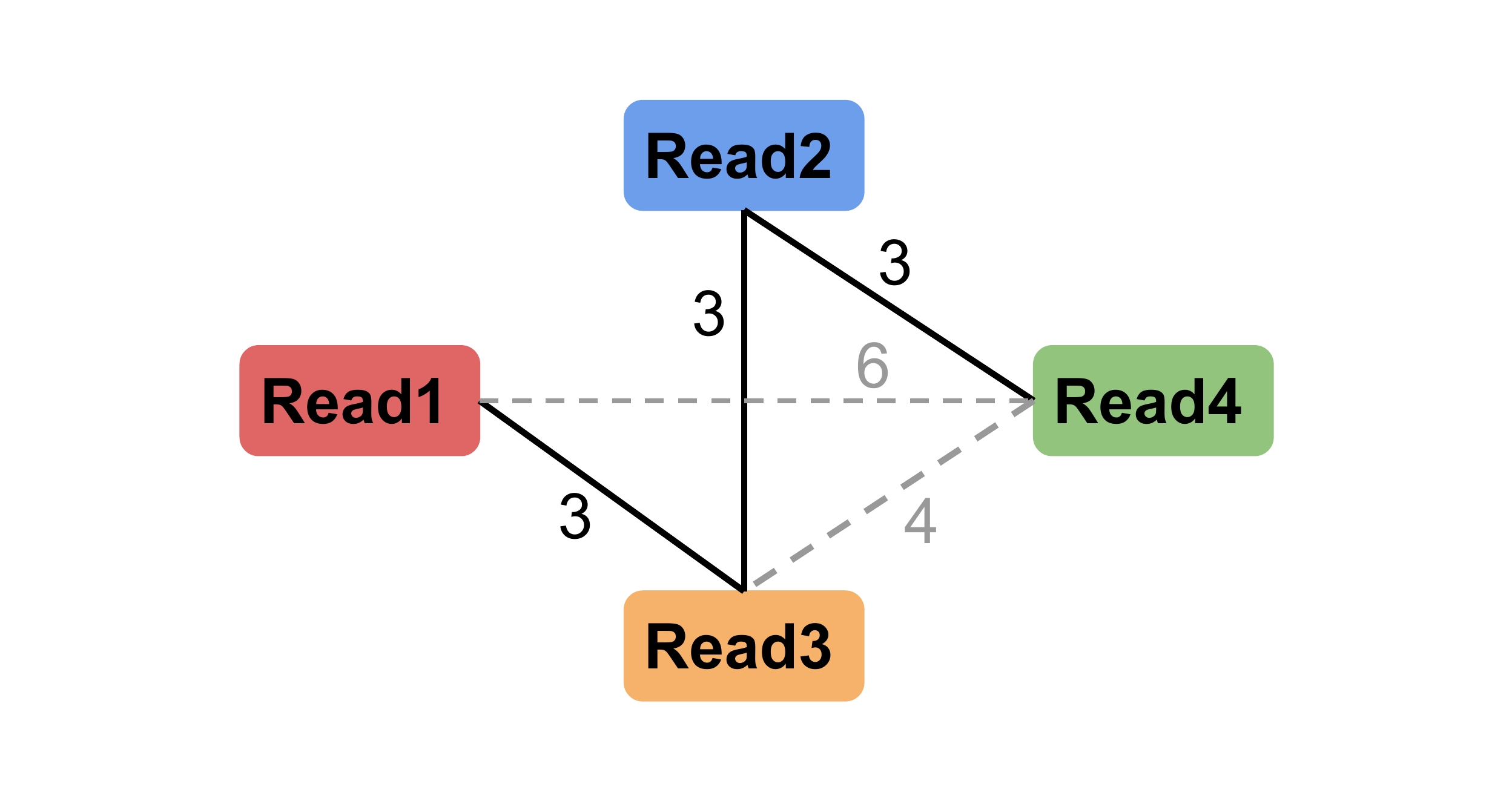}
			}			
		}		
	\end{center}
	
	\caption{
		\label{fig:representative_strategy}
		Example of representative $k$-mer strategy work for four reads of various length, $\tau = 3$ and $k = 2$.\newline
		(Upper left) Multiple alignment for input reads. 
		Columns corresponding to mismatches are highlighted in bold.\newline
		(Upper right) To construct HG algorithm computes representative $k$-mers algorithm using $k$-mer index that
		for each $k$-mer from input reads stores indexes of reads sharing this $k$-mer.
		For example, \textsc{AC} is presented in $Read_1$ only, while \textsc{GT} $2$-mer is presented in all $4$ reads.
		For a given $k$-mer its \textit{multiplicity} is a number of reads containing this $k$-mer.
		For each read algorithm finds representative $k$-mers minimizing a sum of their multiplicities.
		For example, for blue $Read_2$ \textsc{GG}, \textsc{TC}, \textsc{AG}, and \textsc{CT} $2$-mers (labeled by blue square in $k$-mer index) provide the minimal sum of multiplicities that is equal to $8$.\newline
		(Lower left) To compute pairs of reads that will be edges in HG algorithm finds $\tau + 1$ non overlapping representative $2$-mers for each read.
		Representative $2$-mers are highlighted in read color.
		For example, for red $Read_1$ \textsc{CG}, \textsc{AC}, \textsc{AC}, and \textsc{AT} representative $2$-mers were computed (note one $2$-mer can be selected more than one times).
		Occurrence of representative $k$-mers in other reads are shown by lines of color corresponding to read.
		For example, blue $Read_2$ contains representative $2$-mers of orange $Read_3$ (\textsc{GC}) and green $Read_4$ (\textsc{AG}), but does not contain representative $2$-mers of red $Read_1$.
		Thus, algorithm computes $\mathcal{HD}$ only for pairs of reads sharing at least one representative $k$-mer.
		In our case, $\mathcal{HD}$ will not be computed for pair $(Read_1, Read_2)$.\newline
		(Lower right) The resulting Hamming graph consists of three edges: $(Read_1, Read_3)$, $(Read_2, Read_3)$, and $(Read_2, Read_4)$.
		Weights of all edges are equal to $3$.
		Grey dashed edges correspond to pairs for which $\mathcal{HD}$ was computed, but exceeds threshold $\tau$.}
\end{figure*}

\paragraph{Representative $k$-mers strategy.}
To address this problem we proposed \emph{representative $k$-mer strategy} (Algorithm~\ref{alg:hgraphconstructor}).
Algorithm uses the following observation.
If Hamming distance between strings $s_1$ and $s_2$ does not exceed $\tau$,
then each set of $\tau+1$ non overlapping substrings of $s_1$ contains at least one substring of $s_2$.
Note that these substrings do not necessarily cover entire $s_1$ string.
Proposed strategy uses $k$-mers as substrings and for string~$s$ selects the most rare $k$-mers to
find candidates to be connected by edge with $s$.
Note that even in case of short $k$-mers strategy provides an exact solution of HG construction problem. 
To find the most distinguishing $k$-mers algorithm constructs \textsc{KmerIndex} (map from $k$-mer into a set of reads sharing it).
\textit{Multiplicity} of $k$-mer is a number of reads containing this $k$-mer.
Then for each read algorithm finds $\tau+1$ non overlapping $k$-mers minimizing sum of their multiplicities (\textsc{ReprKmers}).
Note that total multiplicity is exactly the number of $\mathcal{HD}$ computations for the current read.
Selected $k$-mers are called \emph{representative} and can be found using dynamic programming algorithm (Algorithm~1 in Appendix~B).
As a result, Hamming distance is computed for all pairs of reads sharing at least one representative $k$-mer. 
Fig.~\ref{fig:representative_strategy} illustrates example of representative $k$-mer strategy work for $4$ strings and $\tau = 3$.
\begin{algorithm}
\caption{\hgconstructor workflow}\label{alg:hgraphconstructor}
\begin{algorithmic}[1]
\Procedure{\hgconstructor}{$\Reads, k, \tau$}
  \State \textsc{KmerIndex} $\gets \varnothing$

  \For {$\Read \in \Reads$}
    \For {$\kappa \in \Call{$k$-mers}{\text{\Read}}$}
      \State \textsc{KmerIndex}($\kappa$) $\gets$ $\Read$
    \EndFor
  \EndFor

  \State \textsc{Graph} $\gets \varnothing$

  \For {$\Read_1 \in \Reads$}
    \State $\Kmers \gets \Call{$k$-mers}{\Read_1}$
    \For {$\kappa \in \Call{ReprKmers}{\Kmers,\tau+1}$}
      \For {$\Read_2 \in \textsc{KmerIndex}(\kappa)$}
        \If {$\mathcal{HD}(\Reads_1, \Reads_2) \leq \tau$}
          \State \textsc{Graph} $\gets$ $\textsc{Graph} \cup \{(\Read_1, \Read_2)\}$
        \EndIf
      \EndFor
    \EndFor
  \EndFor

  \State \Return \textsc{Graph}
\EndProcedure
\end{algorithmic}
\end{algorithm}

\begin{figure*}
	\begin{center}
	\mbox{
		\subfigure[]{
			\label{fig:hg_complexity}
			\includegraphics[width = .45\textwidth]{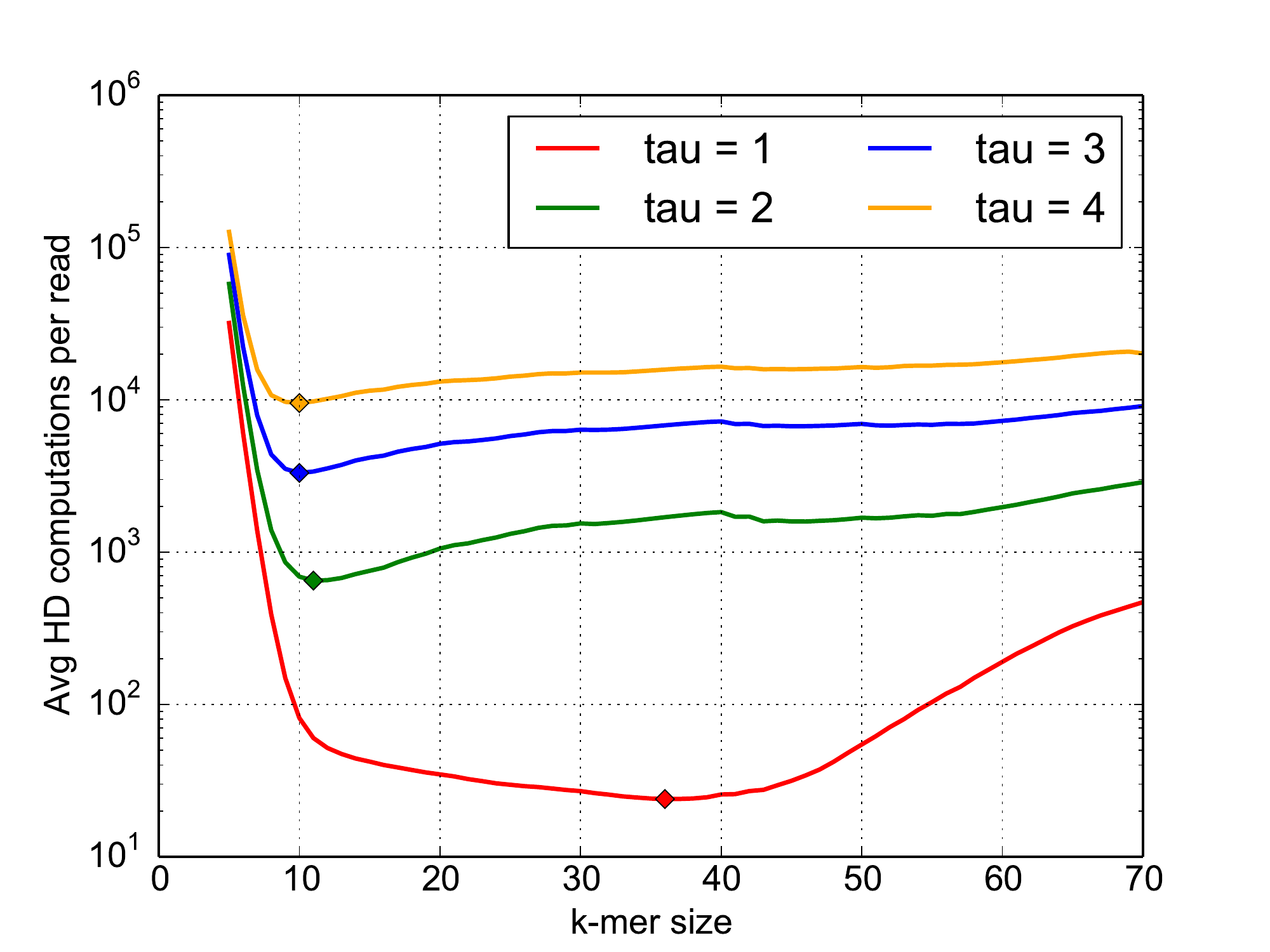}
		}
		\subfigure[]{
			\label{fig:representative_kmer_pos_hist}
			\includegraphics[width = .45\textwidth]{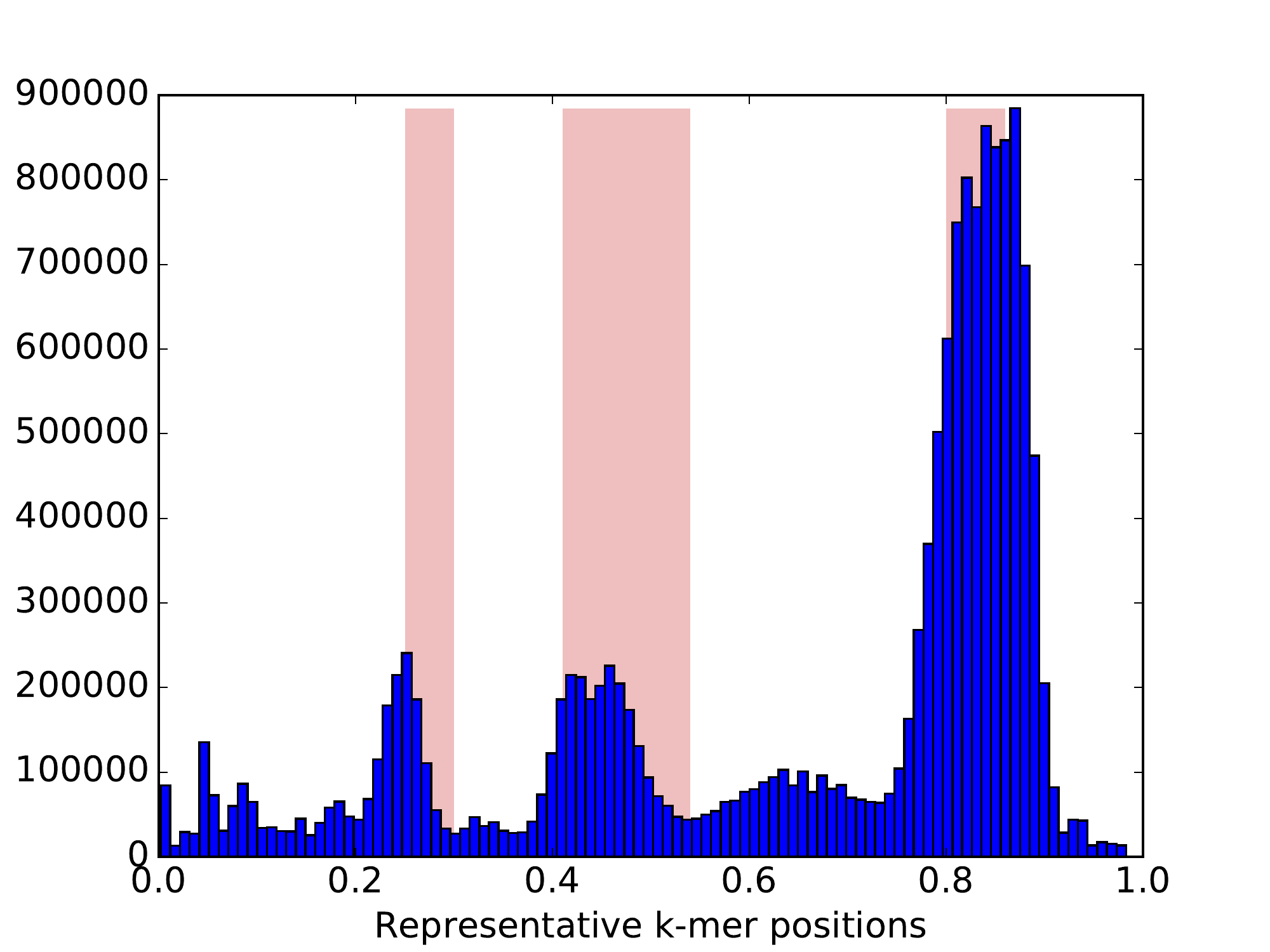}
		}
	}
	\end{center}
	\caption{
		\label{fig:hg_stats}
		(Left) Average number of $\mathcal{HD}$ computations per read for $\tau = 1, 2, 3$ and $4$.
		The best $k$ value for $\tau = 1$ is $35$.
		The best $k$ value for other $\tau$ values is close to $10$.
		(Right) Histogram shows distribution of relative positions of representative $k$-mer for $k = 10$ and $\tau = 4$ for immunosequencing data of heavy chain repertoire.
		Red bars correspond to relative positions of complementary determining regions (CDRs) that are the most diverged regions of immunoglobulins.
		Plot shows that proposed strategy selects $k$-mers from the most representative regions of antibody sequences.
	}
\end{figure*}

\paragraph{\hgconstructor discussion.}
We proposed a novel algorithm for construction of Hamming graph.
The proposed algorithm is applicable for strings of non equal length and can be extended for edit distance (e.g., for correction insertions and deletions introduced at amplification stage).
To reduce number of $\mathcal{HD}$ computations algorithm selects pairs of strings sharing the most distinguishing parts.
However, algorithm performance depends on choice of $k$ value.
Small value of $k$ may lead to a large number of random hits and extra computations of $\mathcal{HD}$.
On the other hand, positions of large non overlapping $k$-mers are almost fixed in the sequence and their total multiplicities can only be changed within a small neighborhood.
For example, if $k=L / (\tau + 1)$ the only one decomposition into $\tau + 1$ non overlapping $k$-mers is possible for a read of length $L$.
Fig.~\ref{fig:hg_complexity} shows plot of average number of $\mathcal{HD}$ computations per read
depending on $k$-mer size on heavy chain repertoire for $\tau = 1, 2, 3$, and $4$.
$k = 35$ provides the optimal number of $\mathcal{HD}$ computations for $\tau = 1$.
Surprisingly, $k$ close to $10$ shows the best results for $\tau = 2, 3$ and $4$.
By default, \igrc uses $k = 10$ and $\tau = 4$ for HG construction.
Fig.~\ref{fig:representative_kmer_pos_hist} shows histogram of positions distribution of representative $k$-mers for $k = 10$ and $\tau = 4$.
Histogram shows that proposed strategy selects $k$-mers from complementary determining regions (CDRs) that are the most distinguishing parts of immunoglobulins.
We tested proposed algorithms on other types of heterogeneous sequences (example on \emph{Alu} repeats is given in Appendix~C) and
showed that our algorithm takes into account their features and detects groups of highly similar sequences.

\section{Results}
Benchmarking of repertoire construction tools is an extremely challenging problem due to lack of golden standard test datasets with known reference repertoire.
In 2015 \cite{igsimulator2015} released \igsimulator tool that allows one to simulate repertoire and immunosequencing library corresponding to it.
In this section we benchmark \igrc against \mixcr and \imseq on simulated datasets.
To estimate \igrc accuracy on real data we propose a reference-free method
for analysis of repertoire construction results on immunosequencing data from sorted naive,
antibody secreting and plasma mouse B-cells.
Selected datasets have different diversity levels that allows one to estimate accuracy of repertoire construction tool.

\subsection{Analysis of \igrc on immunosequencing data of sorted B-cells}
We tested \igrc on three sorted mouse datasets: naive B-cells (NCs), antibody secreting cells (ASCs) and plasma cells (PCs) (\cite{greiff_2014_12727}).
NCs are characterized by high diversity of recombination events, low level of antibody expression and absence of somatic hypermutations.
Thus, we expect that repertoire of NCs will consist of small and non similar antibody clusters.
After successful binding with antigen, naive B-cell undergoes the processes of the secondary diversification: clonal expansion and somatic hypermutagenesis.
As a result of the secondary diversification, naive B-cell turns into ASC or plasma cell that can be presented in multiple copies. 
ASCs and PCs form a collection of families that originate from the same B-cell.
Thus, ASCs and PCs are characterized by low diversity of recombination events and high level of somatic hypermutations.
Also PCs have extremely high expression level while ASCs deliver intermediate values.
In this case, we expect that repertoire of PCs will consist of highly abundant and similar clusters.
Characteristics of sorted cells are summarized in Table~\ref{tbl:sorted_cells_features}.
These specific features allow us to estimate sensitivity and specificity of repertoire construction tools.
Particularly, very soft error correction will result in presence of highly similar antibodies in repertoire of NCs,
while aggressive parameters of error correction will remove natural variations in repertoires of ASCs and PCs.

To estimate accuracy of \igrc we perform the following analysis.
First, we construct repertoires for all three samples.
Quantitative characteristics of the constructed clusters show expression levels of cells.
Then, we extract CDR3 sequences for constructed clusters using \migec tool (\cite{Shugay2014}) and group together identical CDR3s.
CDR3 is the most divergent part of antibody since it covers VDJ junction.
Thus, set of all CDRs partially reflect diversity of entire repertoire.
For example, for NCs we expect to observe large number of non similar CDR3s corresponding to non related VDJ recombinations.
For ASCs and PCs we expect to observe large groups of similar CDR3s that came from the same antibody family and differ by SHMs.
To estimate diversity of repertoires we construct Hamming graphs on CDR3s with $\tau = 3$ and analyze structure of large connected components.
 
Table~\ref{tbl:sorted_cells} shows results of such analysis on NCs, ASCs and PCs.
The first part of Table~\ref{tbl:sorted_cells} shows quantitative characteristics of the constructed clusters
(number of clusters and their multiplicities).
As we expected, repertoire of NCs consists of small (size of the largest cluster is $9$) clusters
while repertoires of ASCs and PCs are formed by large clusters.
The sizes of the largest clusters for ASCs and PCs are $400$ and $1093$, respectively.
The second part of Table~\ref{tbl:sorted_cells} shows number of distinct CDR3s and maximal abundance.
NCs are presented by a large number ($\num{39918}$) of distinct CDR3 sequences with small abundances (max abundance is $20$).
ASCs and PCs repertoires have CDR3s with high abundances ($573$ and $2489$, respectively).
This proves that ASCs and PCs repertoires are presented by families originated from the same B-cell and, thus, a single recombination.
The third part of Table \ref{tbl:sorted_cells} shows characteristics of Hamming graphs on CDR3 sequences.
Particularly, the largest connected components corresponding to NCs is sparse ($137$ vertices and $688$ edges).
The largest connected components corresponding to ASCs and PCs are almost dense (edge fill-ins are $0.75$ and $0.33$).

\begin{figure*}
	\begin{center}
	\mbox{
		\subfigure[]{
			\includegraphics[width = .27\textwidth]{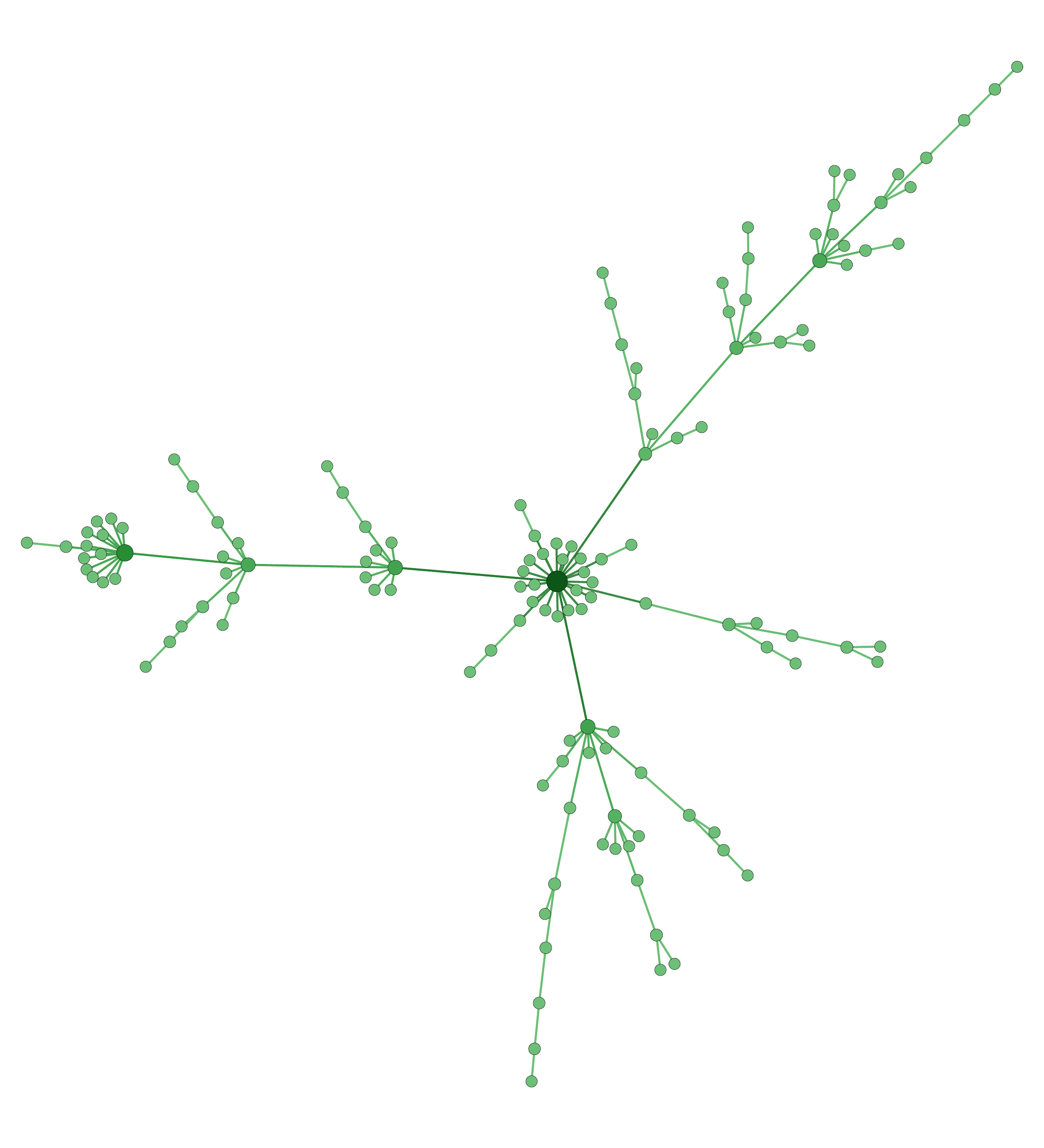}
			\label{fig:naive_mst}
		}
		\hspace{.03\textwidth}
		\subfigure[]{
			\includegraphics[width = .29\textwidth]{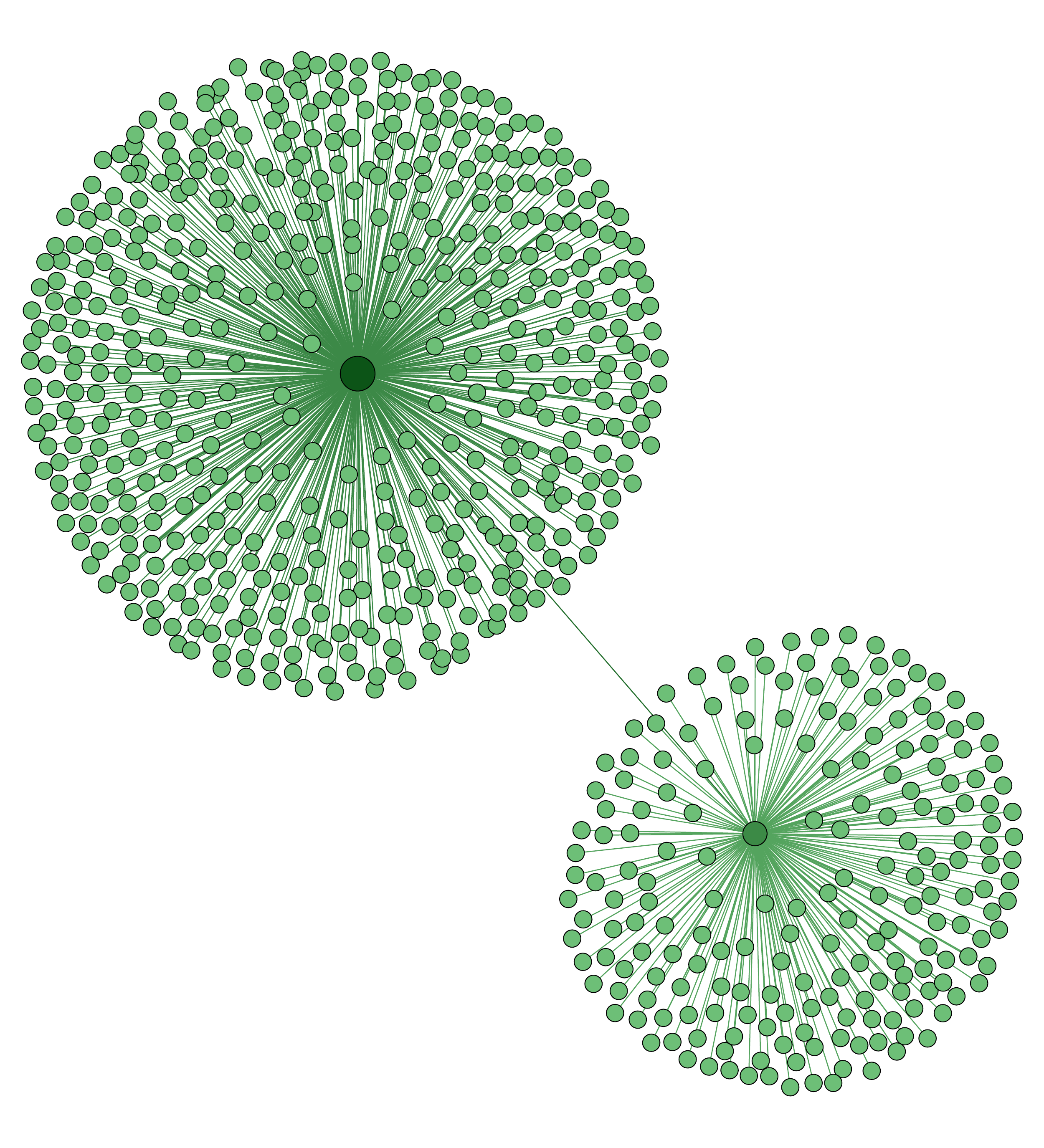}
			\label{fig:asc_mst}
		}
		\hspace{.05\textwidth}
		\subfigure[]{
			\includegraphics[width = .26\textwidth]{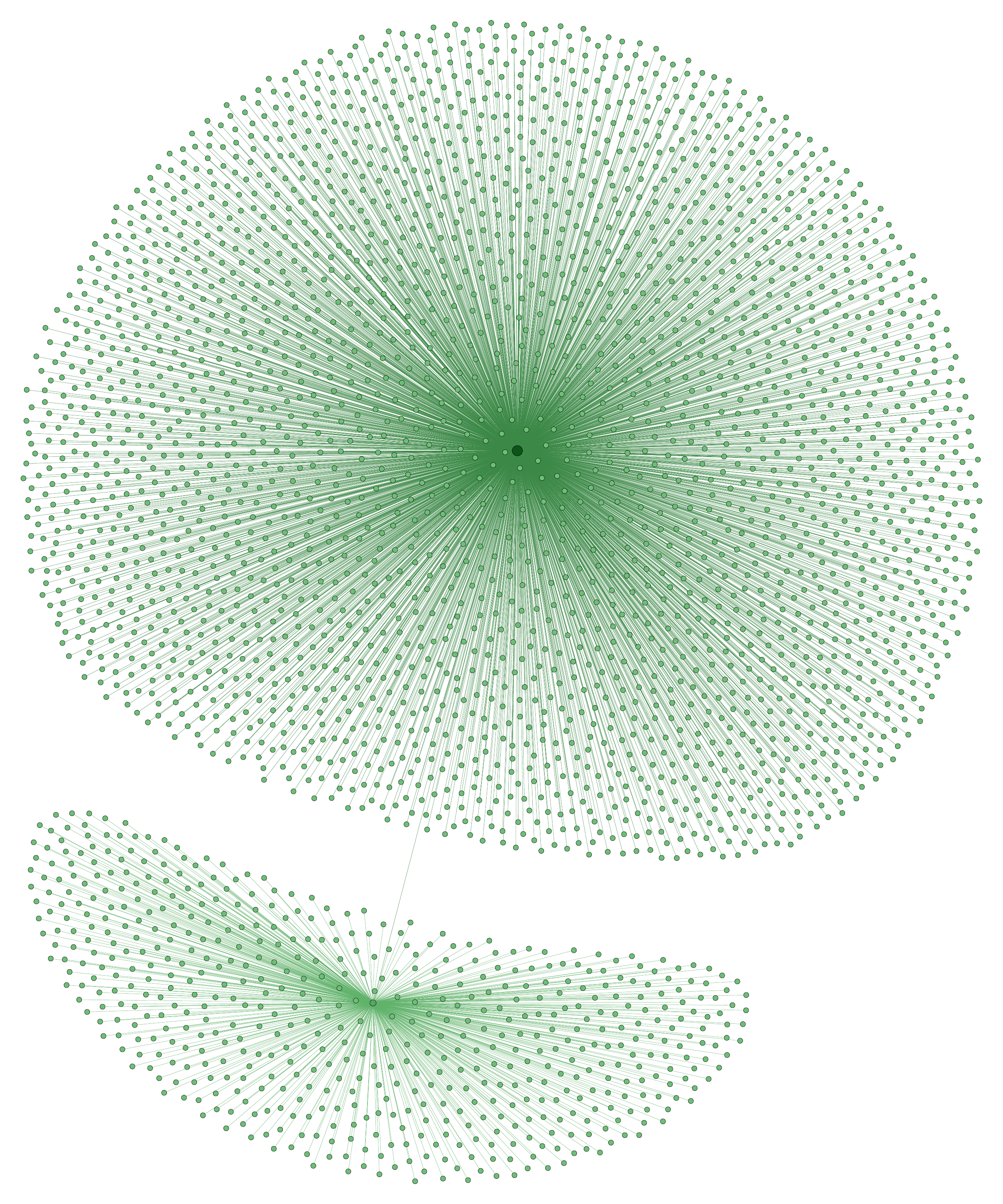}
			\label{fig:pc_mst}
		}				
	}
	\end{center}
	\caption{
	\label{fig:sorted_cells}
	Minimum spanning trees for the largest connected components of Hamming graphs constructed on CDR3 of naive cells (left), ASCs (middle) and plasma cells (right).
	Minimum spanning tree of size $137$ corresponding to naive cells has long branches and presents randomly connected CDR3s.
	In contrary, minimum spanning trees corresponding to ASCs (number of vertices is $702$) and plasma cells (number of vertices is $3213$) look like "stars" (several clones with multiple descendants) and present closely related families of cells.
	}
\end{figure*}

Fig.~\ref{fig:sorted_cells} shows minimum spanning trees for the largest connected components of Hamming graphs on CDR3 for all three samples. 
Table~\ref{tbl:sorted_cells} and Fig.~\ref{fig:sorted_cells} show that \igrc reflects expected properties of sorted cells.
For example, minimum spanning tree corresponding to NCs has long branches and presents randomly connected CDR3s.
In contrast, minimum spanning trees corresponding to ASCs and PCs look like "stars" (few clones with multiple descendants) and present closely related families of cells.

\begin{table*}
	\begin{tabular}{l|c|c|c}
			& 		\textbf{Naive cells (NCs)} & 	\textbf{Antibody secreting cells (ASCs)} & 		\textbf{Plasma cells (PCs)} \\ \hline
		\textit{Recombination diversity} & High & Low & Very low \\
		\textit{Expression level} & Very low & Medium & High \\
		\textit{Secondary diversification} & Does not affect & Affects & Affects \\
		\textit{SHM level} & $-$ & Medium & High \\ 
		\textit{Repertoire features} & Lowly abundant and non-similar clusters & Abundant antibody families & Highly abundant antibody families \\
	\end{tabular}

	\caption{\label{tbl:sorted_cells_features}
		Characteristics of different types of B-cells: naive cells (NCs), antibody secreting cells (ASCs) and plasma cells (PCs).
	}
\end{table*}

\begin{table}
	\begin{tabular}{l|c|c|c}
			& 		\textbf{NCs} & 	\textbf{ASCs} & 		\textbf{PCs} \\ \hline
		\textit{\# reads} & 							$\num{322760}$ & 		$\num{258422}$ & 		$\num{376034}$ \\
		\textit{\# clusters} & 						$\num{231610}$ & 		$\num{112215}$ & 		$\num{131503}$ \\
		\textit{\# non-trivial clusters} & 		$169$ & 							$3115$ & 						$7919$ \\
		\textit{max cluster} & 					$9$ & 								$400$ & 							$1093$ \\ \hline 
		
		\textit{\# CDR3s} & 						$\num{39918}$ & 			$6394$ & 						$6385$ \\
		\textit{max CDR3 abundance} & 		$20$ & 							$573$ & 							$2489$ \\ \hline 
		
		\textit{\# CDR3 components} & 		$1885$ & 						$1038$ & 						$700$ \\
		\textit{max CDR3 component} & 	$137$ & 							$702$ & 							$3213$ \\
		\textit{edge fill-in of max component} & $0.07$ &							$0.75$ &							$0.33$ \\
		\textit{\# trivial CDR3 components} & $\num{37583}$ & 	$9693$ & 						$6388$ \\
	\end{tabular}
	
	\caption{\label{tbl:sorted_cells}
		Analysis of \igrc results on sorted mouse B-cells: naive cells (NCs), antibody secreting cells (ASCs) and plasma cells (PCs).
		The first part of the table shows quantitative characteristics of the constructed repertoires: \textit{\# clusters}, \textit{\# non-trivial clusters}, and \textit{max cluster size}.
		Row \textit{\# non-trivial clusters} shows number of read clusters with multiplicity $>5$.
		The second part of the table shows number of distinct CDR3s and their abundances.
		The third part of the table shows characteristics of Hamming graphs on CDR3 sequences:
		\textit{\# connected components}, \textit{\# trivial components} (i.e., components presented by a single vertex) and
		\textit{size} and \textit{edge fill-in of the largest component}.
	}
\end{table}

\subsection{\igrc benchmarking}

\paragraph{Benchmarking on simulated datasets.}
We benchmarked \igrc against two existing repertoire construction tools: \mixcr (\cite{Bolotin2015}) and \imseq (\cite{Kuchenbecker2015}).
\imseq ignores mutations in V and J gene segments and in fact performs CDR3 classification instead of full-length analysis.
Thus, we excluded \imseq from further benchmarking since CDR3 repertoire cannot be directly compared with full-length repertoire.

To compare \igrc and \mixcr we simulated reference repertoire using \igsimulator tool (\cite{igsimulator2015})\footnote{We launched \igsimulator with the following parameters: number of base sequences $=1000$, expected number of mutated sequences $= \num{10000}$ and expected repertoire size $=\num{50000}$}.
Reference repertoire consists of $\num{13536}$ clusters, $1549$ of them have multiplicity $\geq 5$).
Then we simulated sequencing reads, each sequence from the reference repertoire is covered by one paired-end read on average.
We also introduced sequencing errors following Poisson distribution with various $\lambda \in [0.1, 3.5]$.
As a result, we simulated two types of sequencing libraries: \textit{with answer} (i.e., library contains error-free reads) and \textit{without answer} (i.e., each read in library contains sequencing errors).
To avoid error-free sequences in a library without answer, we introduce one random mismatch in each sequence with zero errors generated by Poisson distribution.
Note that $\lambda$ is equal to the average number of sequencing errors over all reads for the libraries with answer, while for the libraries without answer the average number of errors is $\lambda + e^{-\lambda}$.

For each simulated library we computed two types of metrics: \textit{\% ideal clusters} and \textit{\% extra clusters}.
A cluster from a constructed repertoire is \emph{ideal} if its sequence is exactly presented in the reference repertoire, otherwise such cluster is \emph{extra}.
To find ideal clusters we align all clusters from the constructed repertoire against the reference repertoire and identify perfect
matches (up to short shifts). 
We consider only exact matches since each error in the constructed sequence will impose impact on the further analysis (e.g., analysis of SHMs). Thus, we decided to disregard imperfect matches.
Metric \emph{\% ideal clusters} is a ratio of \emph{\# ideal clusters} to \emph{\# clusters in the reference repertoire.}
Metric \emph{\% extra clusters} is a ratio of \emph{\# extra clusters} to \emph{\# all clusters in the constructed repertoire}.
Note that \emph{\% ideal clusters} and \emph{\% extra clusters} mean sensitivity and false discovery rate, respectively.
Often highly abundant clusters present special interest for further immunological analysis.
Thus, in addition to computing these metrics for all clusters we also computed them for large clusters separately (i.e., clusters with multiplicity $\geq 5$).

\begin{figure*}
	\begin{center}
		\mbox{
			\subfigure[]{
				\label{fig:ideal_with_answer}
				\includegraphics[width = .45\textwidth]{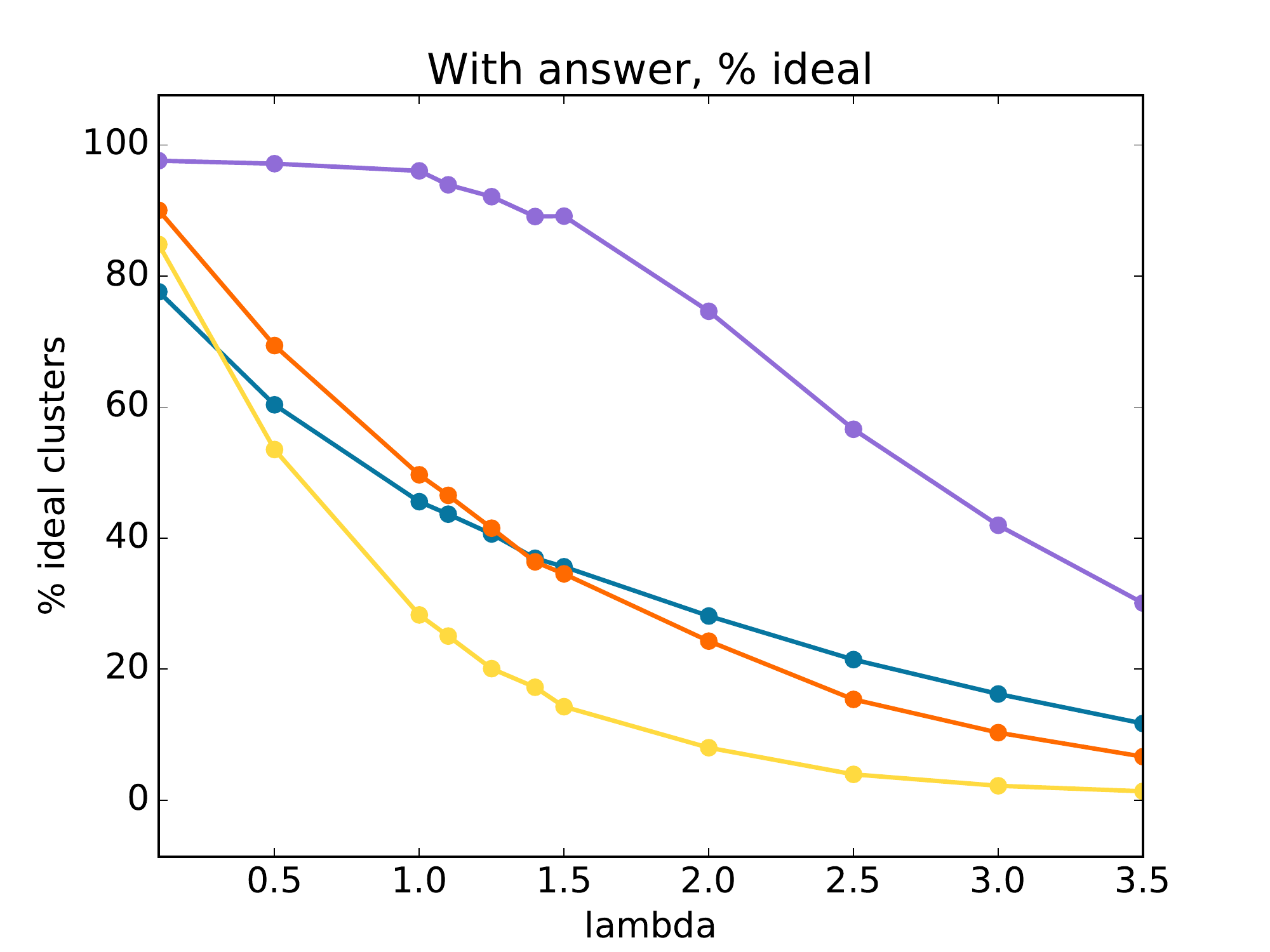}
			}
			\subfigure[]{
				\label{fig:extra_with_answer}			
				\includegraphics[width = .45\textwidth]{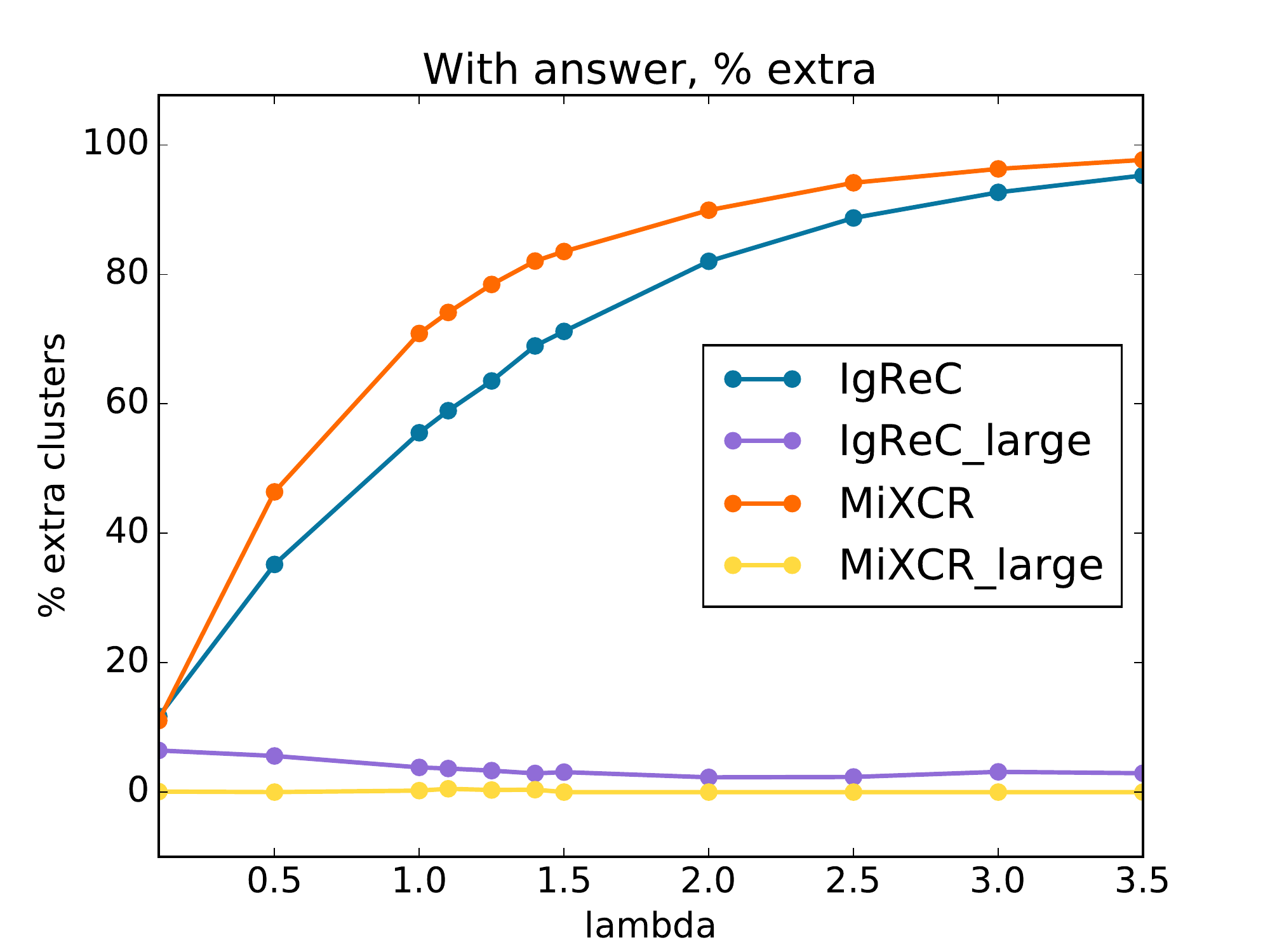}
			}			
		}\\
		\mbox{
			\subfigure[]{
				\label{fig:ideal_without_answer}
				\includegraphics[width = .45\textwidth]{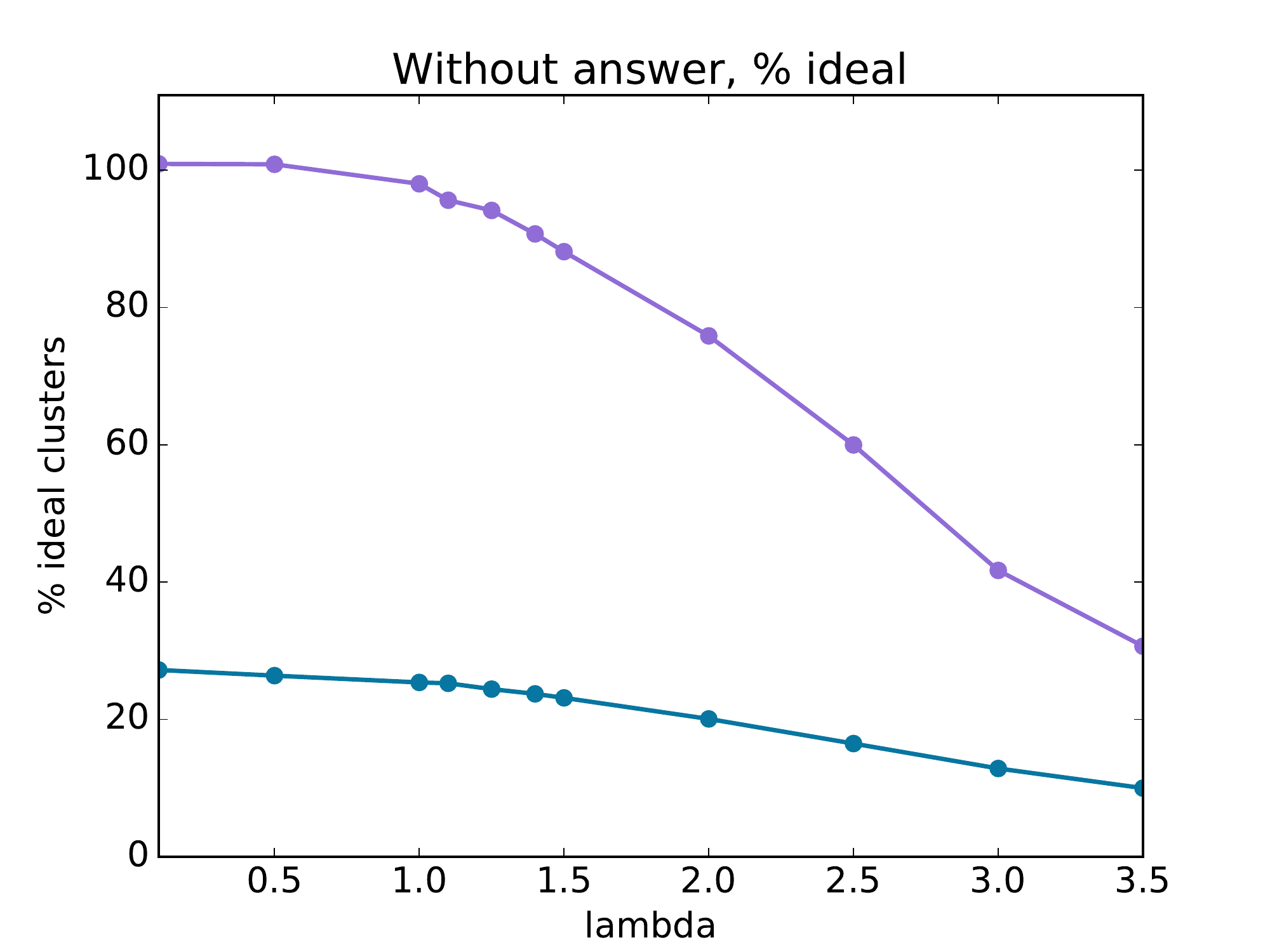}
			}
			\subfigure[]{
				\label{fig:extra_without_answer}
				\includegraphics[width = .45\textwidth]{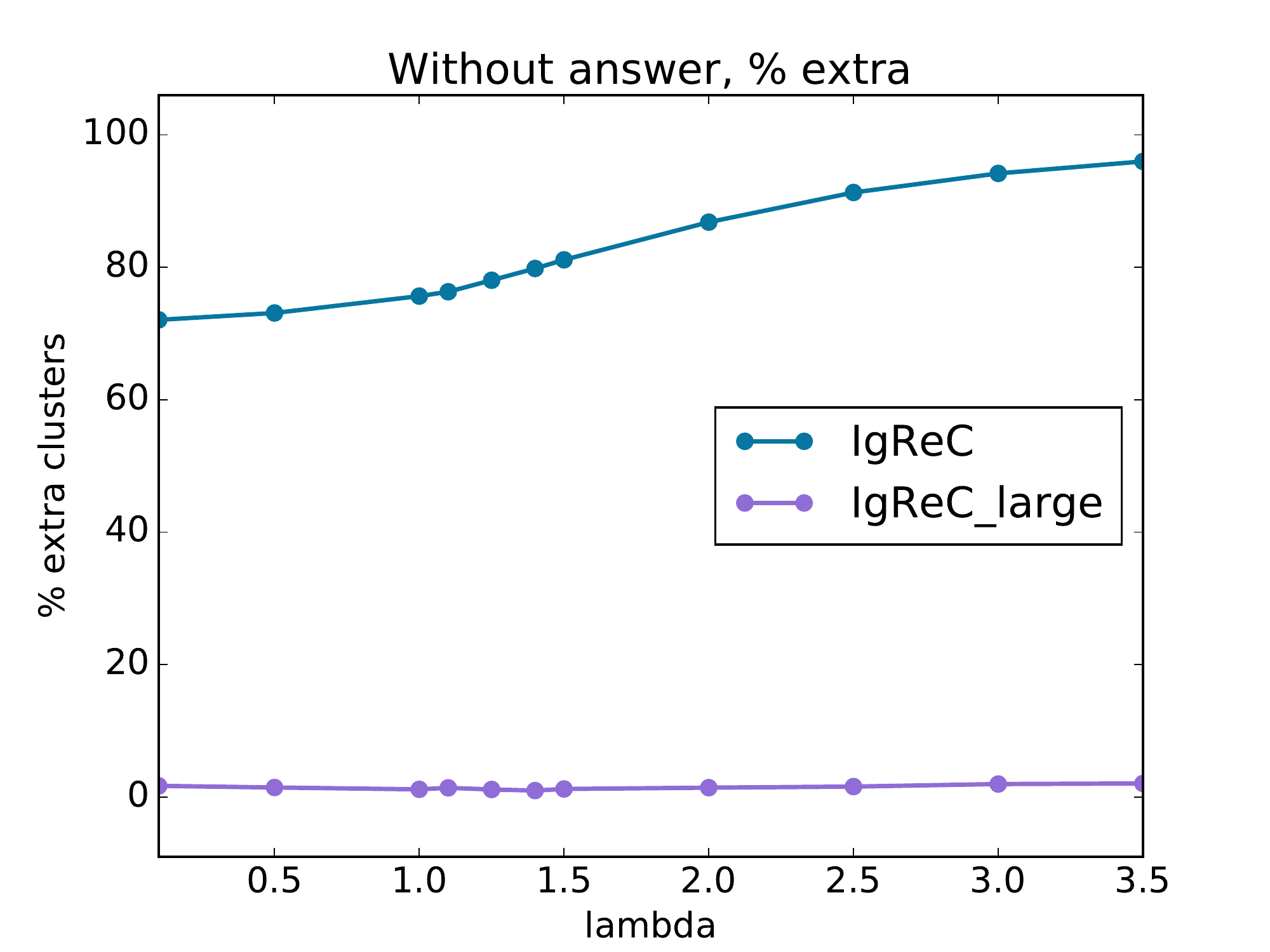}	
				}			
		}		
	\end{center}
	
	\caption{
		\label{fig:igrc_benchmarking}
		(Upper left) and (Upper right) Plots showing \% of ideal and extra clusters constructed by \igrc and \mixcr on libraries with answer.
		(Lower left) and (Lower right) Plots showing \% of ideal and extra clusters constructed by \igrc on libraries without answer.
	}
\end{figure*}

Fig.~\ref{fig:ideal_with_answer} and \ref{fig:extra_with_answer} show metrics for libraries with answer for \igrc and \mixcr repertoires.
\mixcr works fast and produces good results on reads with low error rate ($\lambda < 1.5$).
\mixcr correctly reconstructs most reference clusters for small $\lambda = 0.1$, but quality of its results drops with increasing $\lambda$.
\igrc shows similar results for entire repertoire, but demonstrates an ability to recover large clusters even for high error (average \% of ideal large clusters is $78.04$).
Both \mixcr and \igrc report large number of extra small clusters for $\lambda > 1$, but number of extra large clusters is close to $0$.

Fig.~\ref{fig:ideal_without_answer} and \ref{fig:extra_without_answer} show metrics for libraries without answer.
\mixcr expects that input reads contain well-covered error-free sequences and thus reports very low number of true clusters for libraries without answer.
In contrast, \igrc is able to correctly construct clusters even from highly erroneous reads.
Thus, results of \igrc for libraries without answer are very close to results for libraries with answer (average \% of ideal large clusters is $79.68$).
Also \igrc does not report extra large clusters on libraries without answer, but produces a number of small erroneous clusters.
However, these clusters can be further discarded by multiplicity threshold.

We also performed analysis of multiplicities for the constructed repertoires (Fig.~A2 in Appendix).  
For each ideal cluster we compared its multiplicity in the reference repertoire with its multiplicity in the constructed repertoire. 
Often an ideal reference cluster is broken into several clusters in the constructed repertoire.
Typically these clusters include a large cluster with correct sequence and a number of clusters corresponding to highly erroneous reads.
The larger error rate is the more constructed clusters correspond to a reference ideal cluster (Fig.~A2(a, b)).
Thus, the multiplicities of the clusters in the reference repertoire are higher compared to the multiplicities of the clusters in the constructed repertoire.
However, multiplicities in the reference and constructed repertoires change monotonically: ideal cluster with larger multiplicity in the reference repertoire has larger multiplicity in the constructed repertoire (Fig.~A2(c, d)).

\paragraph{\igrc speedup.}
We compared running time of \igrc with running times of \igrepcon and \mixcr on 6 real human heavy chain datasets (accession numbers in SRA are SRR1383460--5).
These datasets were taken from cervical lymph nodes and contain highly mutated antibody families, i.e., large number of highly similar antibodies.
Thus, such datasets present a challenge for repertoire construction algorithm.
The running time of \igrepcon exceeds $25$ hours on every dataset.
Average running time of \igrc is $\sim 12.35$ minutes.
At the same time, \mixcr works three times faster than \igrc ($\sim 4.23$ min for \mixcr vs $\sim 12.35$ min for \igrc using 16 threads).

\subsection{Discussion}
In this paper we present \igrc, a novel algorithm for adaptive repertoire construction problem.
We propose novel and fast algorithms for alignment of immunosequencing reads against database of germline V and J genes and construction of Hamming graph.
\igrc is capable of handling complex and highly hypermutated heavy chain MiSeq datasets in minutes compared to days required by \igrepcon (\cite{Safonova2015igrc}).
We propose a reference-free analysis of \igrc results using immunosequencing data from sorted B-cells with various diversity and
expression levels: naive cells, antibody-secreting cells and plasma cells.
We show that \igrc takes into account features of input data and accurately recovers expected level of diversity.
Also we benchmark \igrc against \mixcr on repertoires simulated using \igsimulator tool.
We simulate a number of sequencing libraries with various error rates.
\igrc outperforms \mixcr on datasets, where expected number of sequencing errors per read exceeds $1.5$.
At the same time, \mixcr provides better results for low error rate (expected number of sequencing errors per reads is $<1.5$).
Thus, the main advantage of \igrc is high accuracy even in case when true sequences are underrepresented in input reads,
while \mixcr fails to produce comprehensive results in this case.

\igrc tool can be applicable for both antibody and TCR repertoires.
\igrc demonstrates high accuracy compared to other solutions and improves state-of-the-art of adaptive immune repertoire construction problem.

\section*{Acknowledgement}
We are grateful to Anton Bankevich for effective collaboration, insightful comments and help in preparation of this paper.

\paragraph{Funding.}
All authors are supported by Russian Science Foundation (grant No 14-50-00069).


\bibliographystyle{natbib}
\bibliography{igrc2}

\begin{appendices}
\setcounter{figure}{0} \renewcommand{\thefigure}{A\arabic{figure}}
\setcounter{algorithm}{0} \renewcommand{\thealgorithm}{A\arabic{algorithm}}

\section{\vjfinder benchmarking}
We benchmarked \vjfinder on human heavy chain repertoire (accession number SRR1383463) containing $\num{1852661}$ paired reads. 
$\num{1405177}$ reads were successfully merged.
\igblast was chosen as a reference V(D)J labeling tool.

\vjfinder successfully found all
$\num{37435}$ contamination reads (reads with \igblast $e$-value of V gene alignment greater than $0.001$ were considered as contaminations).
\vjfinder also reported $\num{28537}$ reads where V or J genes are not fully presented.
For remaining $\num{1337956}$ reads we compared start positions of V gene segment and end positions of J gene segment reported by \igblast and \vjfinder.
\vjfinder results were inconsistent with \igblast output only for $0.093\%$ of reads.
\vjfinder with default parameters is $20$ times faster than \igblast ($	\sim 6$ mins in $16$ threads by \vjfinder vs $	\sim 30$ hours in single thread by \igblast).

\section{Representative k-mers finding algorithm}
For the sake of simplicity \textsc{ReprKmers} procedure (Algorithm~\ref{alg:representative_kmers}) is defined recursively.
However, one can notice that procedure is performed only for prefixes
$\Kmers[1:j], j = 1,\dots,|\Kmers|$ and for $n = 0,\dots \tau+1$.
One can use dynamic programming: sequentially compute and cache multiplicities of results
and then restore optimal $k$-mer subset.
Thus, overall time and memory complexity is $\mathcal{O}(|\Kmers| \cdot n)$

\begin{algorithm}
\caption{Representative $k$-mer finding algorithm}\label{alg:representative_kmers}
\begin{algorithmic}[1]

\State \textbf{global} $k$, \textsc{KmerIndex}
\Procedure{ReprKmers}{\Kmers, $n$}

  \If {$n = 0$}
    \State \Return $\varnothing$
  \EndIf

  \State \textsc{NumKmers} $\gets$ $|\Kmers|$

  \If {$ \textsc{NumKmers} = 1 + (n-1)k$}
    \State \Return $\Kmers_1, \Kmers_{1 + k}, \dots,  \Kmers_{1 + (n-1)k}$
  \EndIf

  \State $K_{skip} \gets \Call{ReprKmers}{\Kmers[1 : \textsc{NumKmers} - 1], n}$
  \State $K_{take} \gets \Call{ReprKmers}{\Kmers[1 : \textsc{NumKmers}- k], {n-1}},
  \Kmers_\textsc{NumKmers}$

  \If {$\Call{Multiplicity}{K_{skip}} < \Call{Multiplicity}{K_{take}}$}
    \State \Return $K_{skip}$
  \Else
    \State \Return $K_{take}$
  \EndIf
\EndProcedure
\end{algorithmic}
\end{algorithm}


\section{Results of representative k-mer strategy on Alu repeats}
Alu element is the most representative short repeat in mammalian genomes (length of Alu is $\sim 300$ nt).
About $10.7\%$ of the human genome consists of Alu sequences.
Database of Alu repeats includes three main subfamilies: \emph{AluJ}, \emph{AluS}, and \emph{AluY}.
Alu insertions can cause certain inherited human diseases and in various forms of cancer.
This makes evolutionary analysis of newly occurred Alu repeats an important bioinformatics problem.
In contrast to phylogenetic tree construction problem, Alu repeats present clonal tree, i.e., trees with intermediate nodes.

\begin{figure*}[t]
	\begin{center}
		\mbox{
			\subfigure[]{
				\label{fig:alu_jr4}
				\includegraphics[width = .45\textwidth]{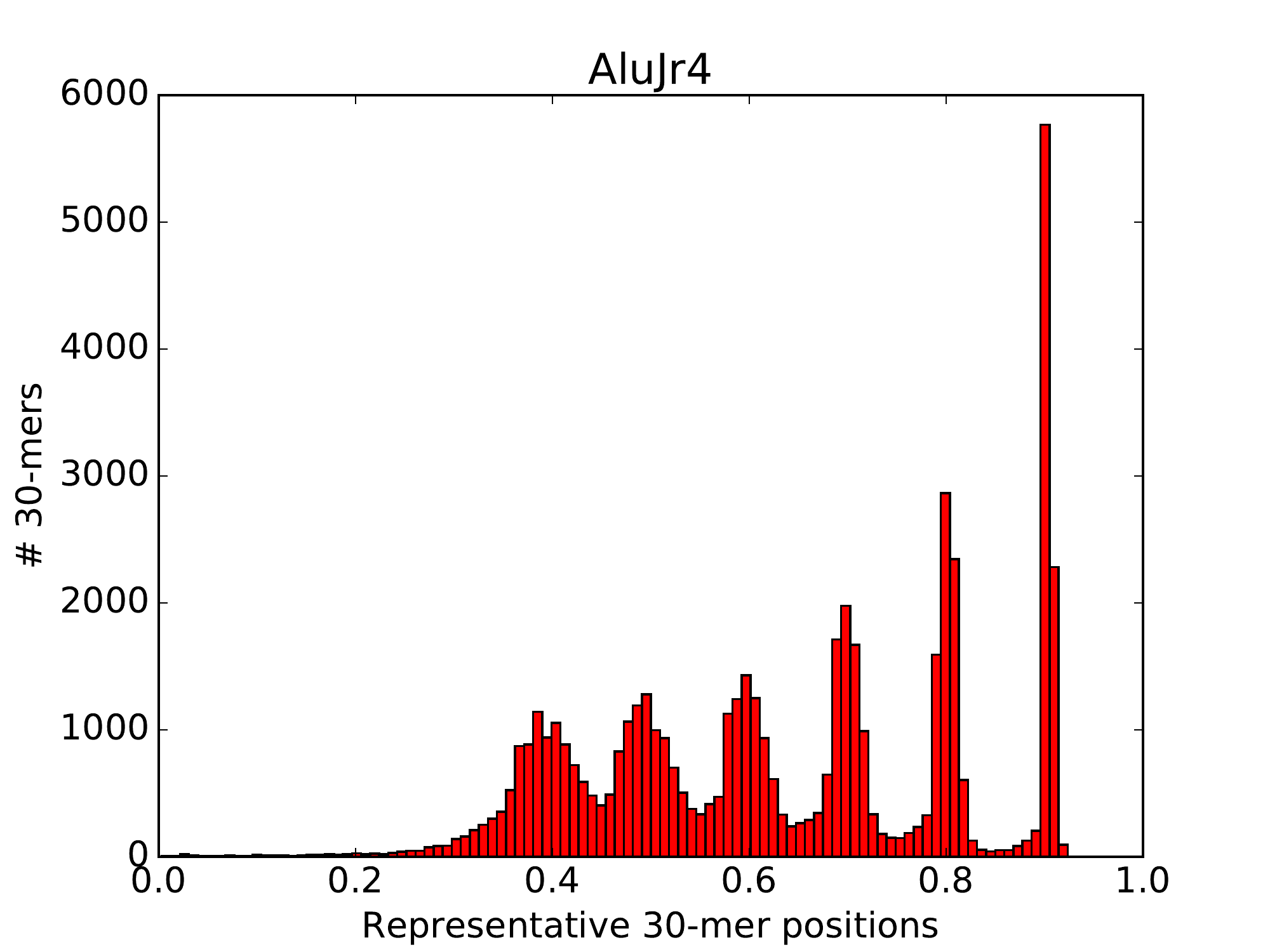}
			}
			\subfigure[]{
				\label{fig:alu_sc5}
				\includegraphics[width = .45\textwidth]{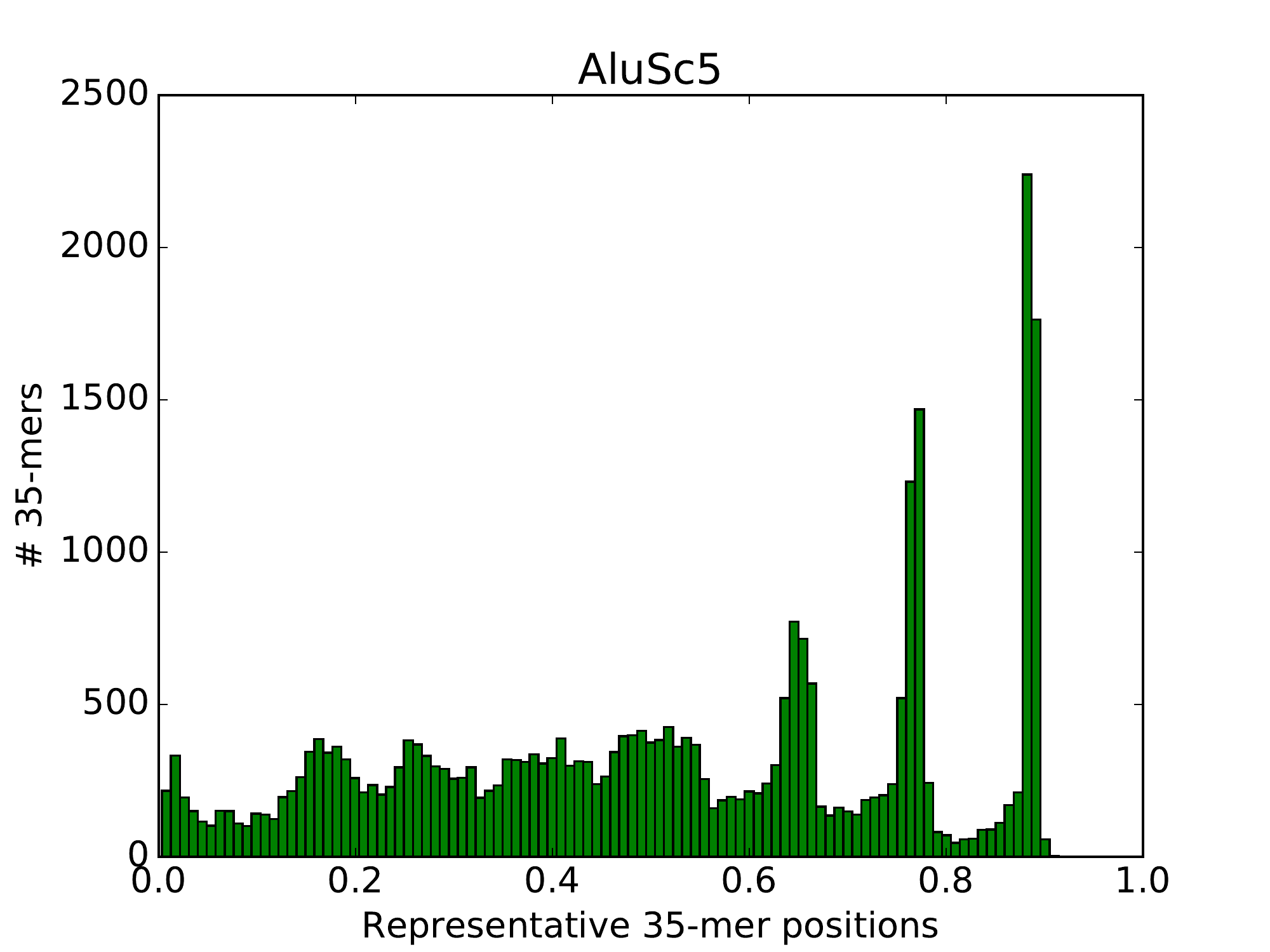}
			}			
		}\\
		\mbox{
			\subfigure[]{
				\label{fig:alu_ym1}
				\includegraphics[width = .45\textwidth]{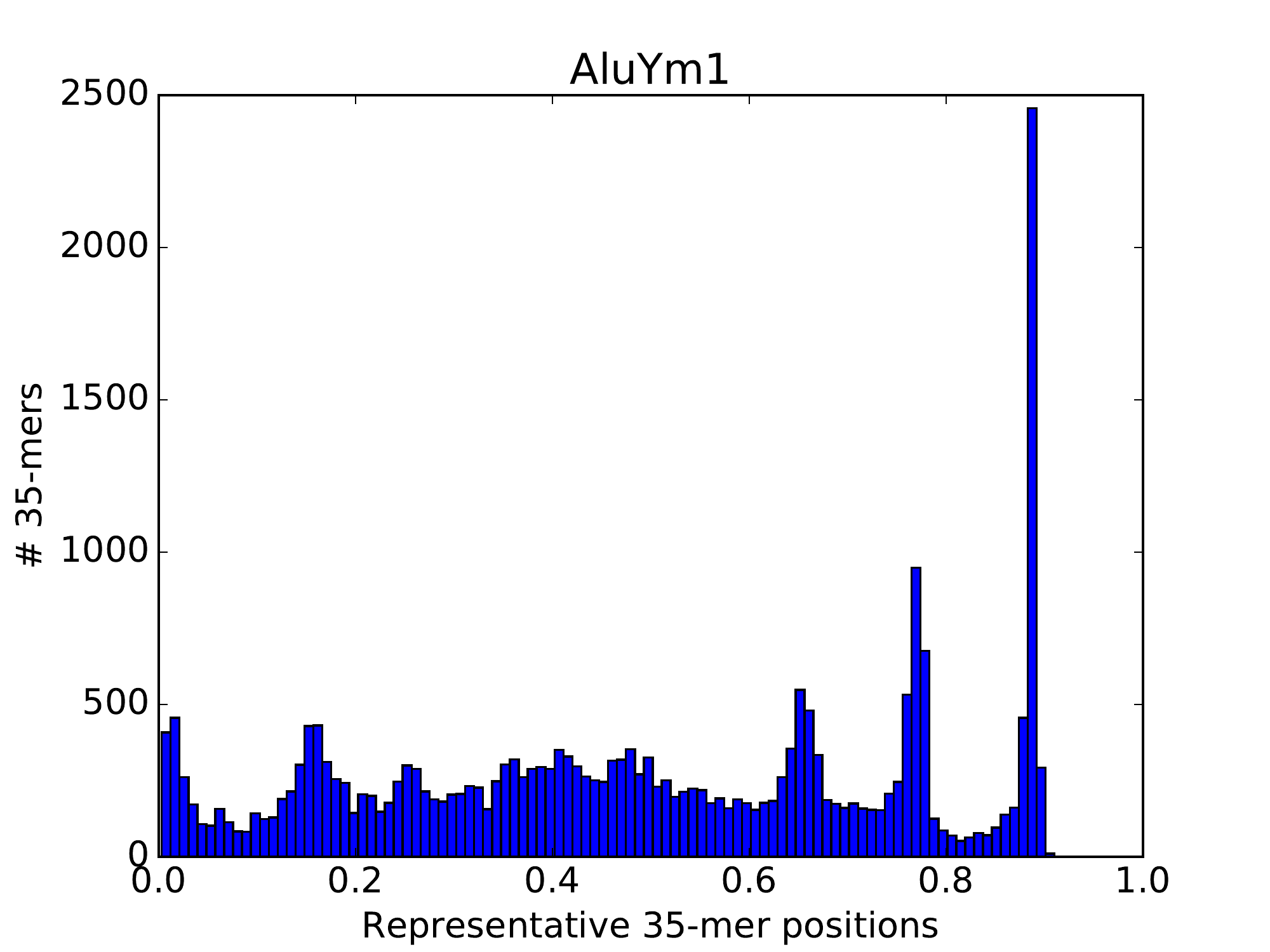}
			}
			\subfigure[]{
				\label{fig:alu_graphs}
                \includegraphics[width = .45\textwidth]{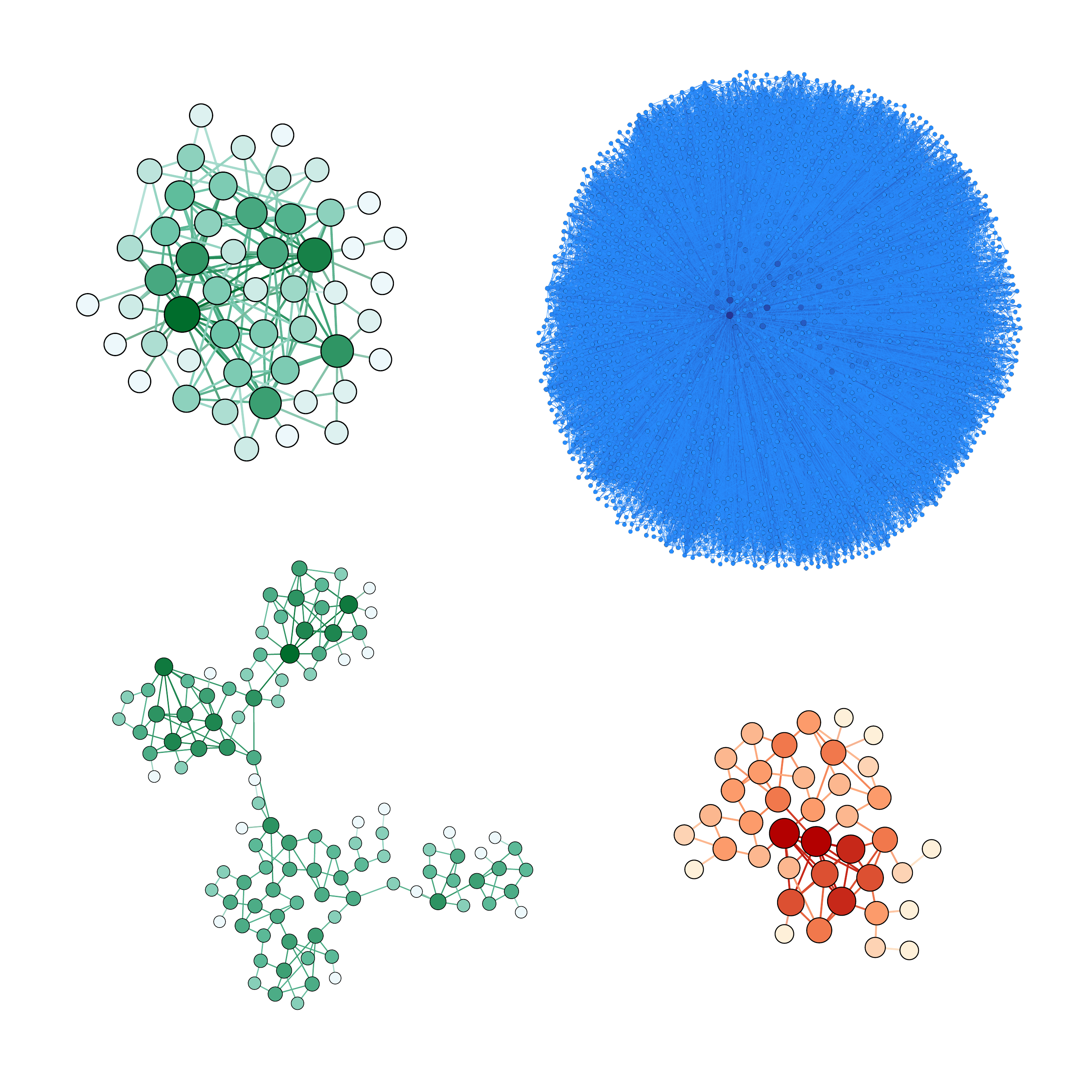}
			}
		}
	\end{center}
	\caption{
		\label{fig:hg_for_alu_repeats}
		\subref{fig:alu_jr4}--\subref{fig:alu_ym1} show histograms of distributions of representative $k$-mer positions for \emph{AluJr4}, \emph{AluSc5} and \emph{AluYm1}, respectively.
    	\subref{fig:alu_graphs} shows four largest connectivity components of HG graph constructed on \emph{AluJr4} (red), \emph{AluSc5} (green) and \emph{AluYm1} (blue).
		Note that all components are one-colored, i.e., represent a single subsubfamily.
	}
\end{figure*}

Alu repeats form a collection of heterogeneous sequences and thus present an excellent test for HG construction algorithm using representative $k$-mers.
We selected three subsubfamilies of Alu repeats: \emph{AluJr4}, \emph{AluSc5} and \emph{AluYm1}.
For each subsubfamily we compute representative $k$-mers for $\tau = 10$.
Fig.~\ref{fig:hg_for_alu_repeats}\subref{fig:alu_jr4}--\subref{fig:alu_ym1} show histograms of distributions of representative $k$-mer positions.
Representative $k$-mers reveal unique features of each subsubfamily.
Particularly, sequences from \emph{AluJr4} have conservative start in contrast to \emph{AluSc5} and \emph{AluYm1}.
Constructed graphs can be efficiently applied for classification of new repeats.
Fig.~\ref{fig:hg_for_alu_repeats}\subref{fig:alu_graphs} shows four largest connectivity components of similarity graph constructed for selected subsubfamilies.
Each connectivity component contains Alu from the same family that confirms an applicability of proposed HG construction approach to repeats classification.

\begin{figure*}[t]
	\begin{center}
		\mbox{
			\subfigure[]{
				\label{fig:m2m_igrec_lam2}
        \includegraphics[width = .45\textwidth]{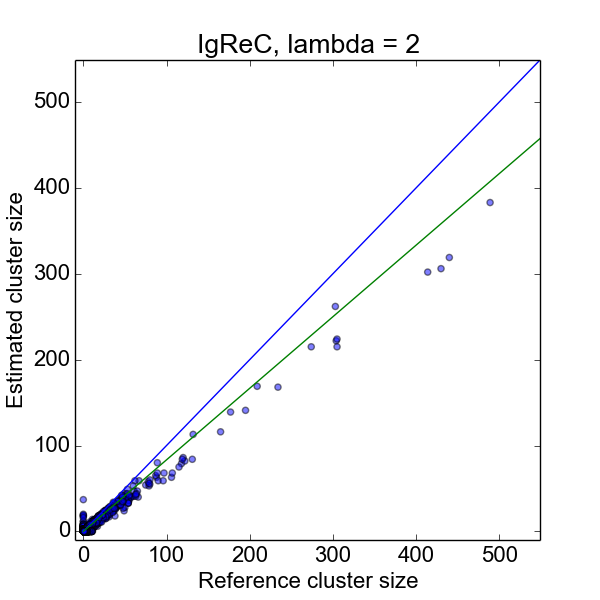}
			}
			\subfigure[]{
				\label{fig:m2m_mixcr_lam2}
        \includegraphics[width = .45\textwidth]{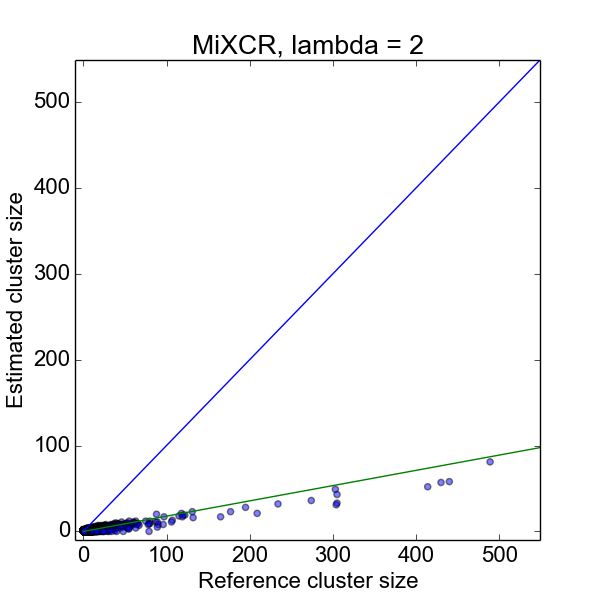}
			}
		}\\
		\mbox{
			\subfigure[]{
				\label{fig:cl_index_w_ans}
        \includegraphics[width = .45\textwidth]{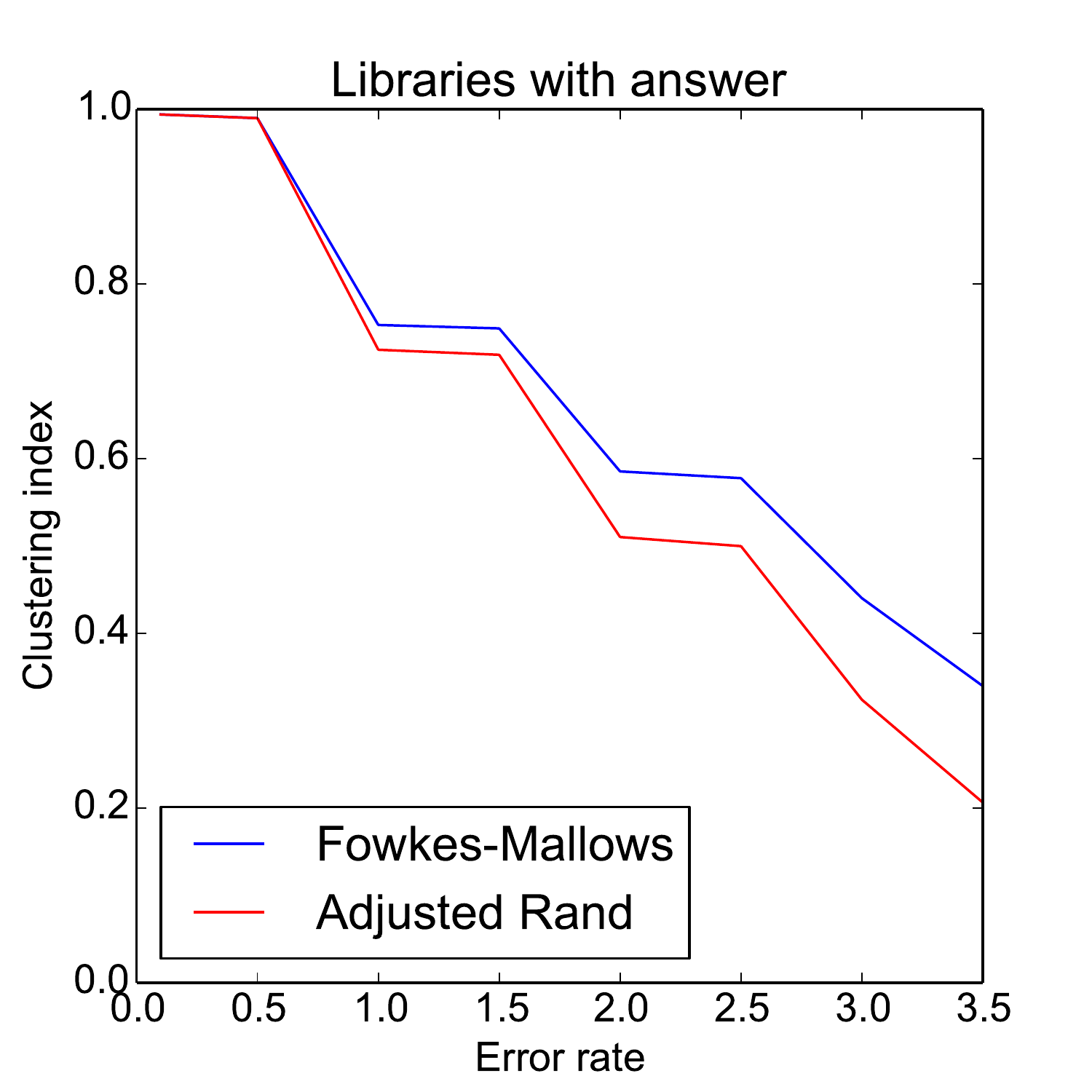}
			}
			\subfigure[]{
				\label{fig:cl_index_wo_ans}
        \includegraphics[width = .45\textwidth]{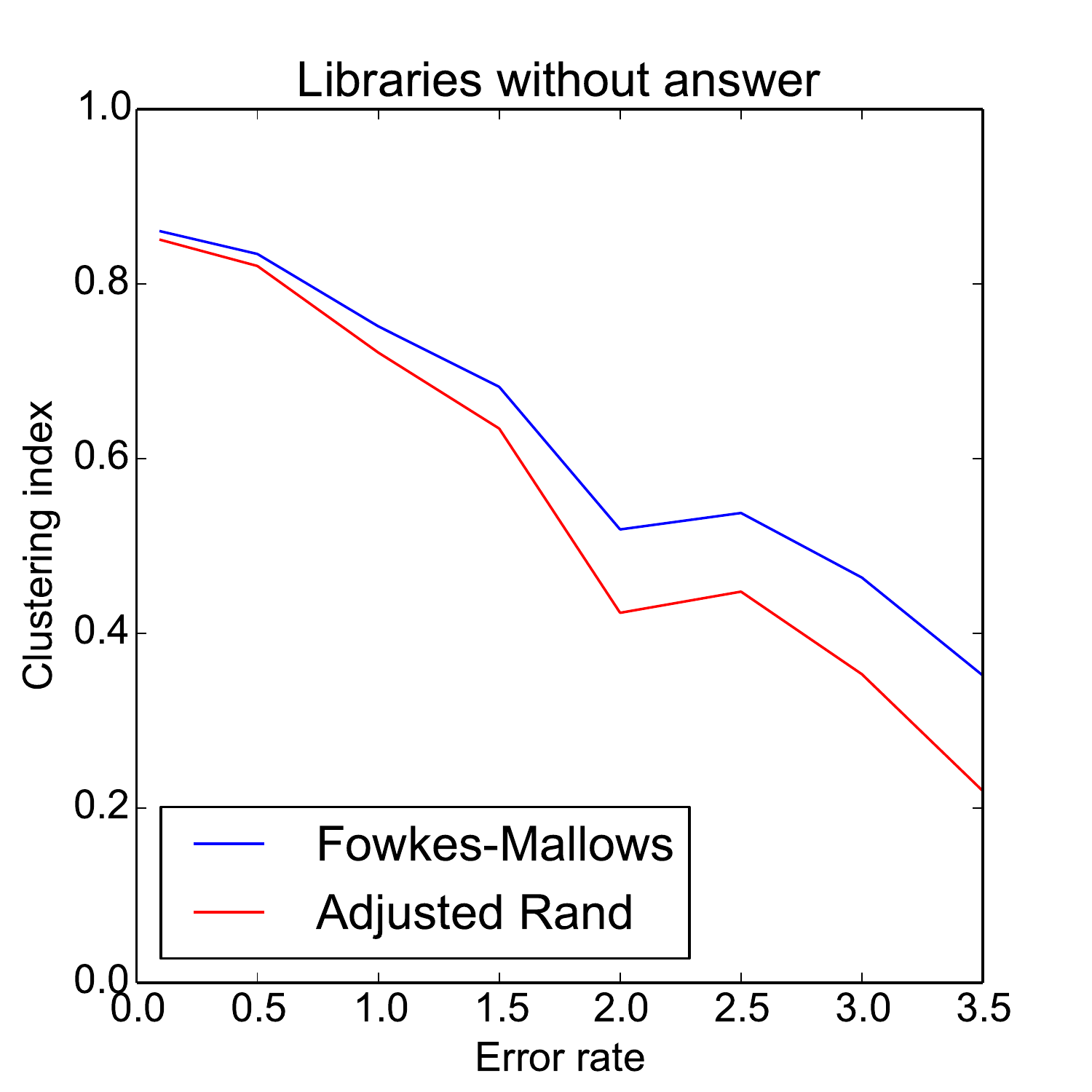}
			}
		}
	\end{center}
	\caption{
		\label{fig:mult_anal}
    (Upper left) and (Upper right) show dependency between multiplicities in reference and repertoire constructed by \igrc and \mixcr, respectively.
    Zero multiplicity means that sequence is not presented in the repertoire. 
    Clearly seen that reconstructed multiplicities are lower than reference ones (blue line) but dependence is almost linear.
    Median ratio (green line) for large clusters (clusters of multiplicity $\geq 5$) for \igrc and \mixcr are $0.83$ and $0.18$, respectively.
    \subref{fig:cl_index_w_ans} and \subref{fig:cl_index_wo_ans} show Fowkes-Mallows and adjusted Rand similarity indices between reference repertoire and repertoire constructed by \igrc for libraries with and without answer, respectively.
	}
\end{figure*}

\end{appendices}
\end{document}